\documentclass[twocolumn]{aastex631}
\usepackage{amsmath}
\usepackage{float}
\usepackage[T1]{fontenc}
\usepackage[utf8]{inputenc}
\usepackage{float}		
\usepackage{amssymb}
\usepackage{textcomp}
\usepackage{gensymb}
\usepackage[procnames]{listings}
\usepackage{tabto}
\usepackage{tabularx}
\usepackage{paralist} 
\usepackage{mathtools}
\usepackage{relsize}
\usepackage[normalem]{ulem}
\usepackage{xcolor}
\usepackage{xspace}
\usepackage[encapsulated]{CJK}
\newcommand{\cntext}[1]{\begin{CJK}{UTF8}{gbsn}#1\end{CJK}}
\usepackage{comment}
\usepackage{ucs}
\usepackage{multirow} 

\defcitealias{Issaoun2022}{Issaoun, Wielgus et al. 2022}

\def\sgra{\object{Sgr~A$^{\ast}$}\xspace}
\def\nrao{\object{NRAO\,530}\xspace}

\graphicspath{{./}{figs/}}

\begin{document}
	
	\title{The Event Horizon Telescope Image of the Quasar NRAO\,530}

	\shorttitle{The EHT Image of the Quasar NRAO\,530}
\shortauthors{Jorstad \& Wielgus et al.}
	
	\email{jorstad@bu.edu}
 \email{maciek.wielgus@gmail.com}

\author[0000-0001-6158-1708]{Svetlana Jorstad}
\affiliation{Institute for Astrophysical Research, Boston University, 725 Commonwealth Ave., Boston, MA 02215, USA}
		
\author[0000-0002-8635-4242]{Maciek Wielgus}
\affiliation{Max-Planck-Institut f\"ur Radioastronomie, Auf dem H\"ugel 69, D-53121 Bonn, Germany}

\author[0000-0001-7361-2460]{Rocco Lico}
\affiliation{Instituto de Astrof\'{\i}sica de Andaluc\'{\i}a-CSIC, Glorieta de la Astronom\'{\i}a s/n, E-18008 Granada, Spain}
\affiliation{INAF-Istituto di Radioastronomia, Via P. Gobetti 101, I-40129 Bologna, Italy}

\author[0000-0002-5297-921X]{Sara Issaoun}
\affiliation{Center for Astrophysics $|$ Harvard \& Smithsonian, 60 Garden Street, Cambridge, MA 02138, USA}
\affiliation{NASA Hubble Fellowship Program, Einstein Fellow}

\author[0000-0002-3351-760X]{Avery E. Broderick}
\affiliation{Perimeter Institute for Theoretical Physics, 31 Caroline Street North, Waterloo, ON, N2L 2Y5, Canada}
\affiliation{Department of Physics and Astronomy, University of Waterloo, 200 University Avenue West, Waterloo, ON, N2L 3G1, Canada}
\affiliation{Waterloo Centre for Astrophysics, University of Waterloo, Waterloo, ON, N2L 3G1, Canada}

\author[0000-0002-5278-9221]{Dominic W. Pesce}
\affiliation{Center for Astrophysics $|$ Harvard \& Smithsonian, 60 Garden Street, Cambridge, MA 02138, USA}
\affiliation{Black Hole Initiative at Harvard University, 20 Garden Street, Cambridge, MA 02138, USA}

\author[0000-0002-7615-7499]{Jun Liu (\cntext{刘俊})}
\affiliation{Max-Planck-Institut f\"ur Radioastronomie, Auf dem H\"ugel 69, D-53121 Bonn, Germany}

\author[0000-0002-4417-1659]{Guang-Yao Zhao}
\affiliation{Instituto de Astrof\'{\i}sica de Andaluc\'{\i}a-CSIC, Glorieta de la Astronom\'{\i}a s/n, E-18008 Granada, Spain}

\author[0000-0002-4892-9586]{Thomas P. Krichbaum}
\affiliation{Max-Planck-Institut f\"ur Radioastronomie, Auf dem H\"ugel 69, D-53121 Bonn, Germany}


\author[0000-0002-9030-642X]{Lindy Blackburn}
\affiliation{Black Hole Initiative at Harvard University, 20 Garden Street, Cambridge, MA 02138, USA}
\affiliation{Center for Astrophysics $|$ Harvard \& Smithsonian, 60 Garden Street, Cambridge, MA 02138, USA}

% ...

\begin{abstract}
We report on the observations of the quasar \nrao with the Event Horizon Telescope (EHT) on 2017 April 5$-$7, when \nrao was used as a calibrator for the EHT observations of Sagittarius~A$^\ast$. At $z=0.902$ this is the most distant object imaged by the EHT so far. We reconstruct the first images of the source at 230\,GHz, at an unprecedented angular resolution of $\sim 20\,\mu$as, both in total intensity and in linear polarization. We do not detect source variability, allowing us to represent the whole data set with static images. The images reveal a bright feature located on the southern end of the jet, which we associate with the core. The feature is linearly polarized, with a fractional polarization of $\sim$5-8\% and has a sub-structure consisting of two components. Their observed brightness temperature suggests that the energy density of the jet is dominated by the magnetic field. The jet extends over 60~$\mu$as along a position angle PA$\sim -$28$^\circ$. It includes two features with orthogonal directions of polarization (electric vector position angle, EVPA), parallel and perpendicular to the jet axis, consistent with a helical structure of the magnetic field in the jet. The outermost feature has a particularly high degree of linear polarization, suggestive of a nearly uniform magnetic field. Future EHT observations will probe the variability of the jet structure on ${\mu}$as scales, while simultaneous multi-wavelength monitoring will provide insight into the high energy emission origin.

~~~~~~~~~~~~~~~~~~~~~~~~~~~~~~~~~~~~~~~~~~~~~~~~~~~~~~~~~~~~~~~~~~~~~~~~
	
\end{abstract}



    \section{Introduction}
    \label{sec:intro} 
\nrao (1730$-$130, J1733$-$1304) is a flat-spectrum radio quasar (FSRQ) that belongs to the class of bright $\gamma$-ray blazars and shows significant variability across the entire electromagnetic spectrum. The source was monitored by the University of Michigan Radio Observatory (UMRAO, USA) at 4.8, 8.4, and 14.5 GHz for several decades\footnote{\url{https://dept.astro.lsa.umich.edu/datasets/umrao.php}} until 2012. The quasar underwent a dramatic radio outburst in 1997  \citep{Aller2009}, during which its flux density at 14.5~GHz exceeded 10 Jy, while the average value is $\sim$2~Jy. Since 2002 \nrao has been monitored by the Sub-Millimeter Array (SMA, Maunakea, Hawai'i) at 1.3~mm and 870~$\mu$m\footnote{\url{http://sma1.sma.hawaii.edu/callist/callist.html?plot=1733-130}}. The data show high amplitude variability at 1.3~mm from 1~Jy to 4 Jy, with the brightest outburst in the end of 2010 contemporaneous with prominent $\gamma$-ray activity \citep{Karen2014}. \cite{Venki2015} have found a statistically significant correlation between the radio and $\gamma$-ray light curves of \nrao, based on measurements at 37~GHz by the Mets\"ahovi Radio Observatory (Aalto University, Finland) and  $\gamma$-ray fluxes measured with the Large Area Telescope (LAT) on board the Fermi Space Gamma-Ray Telescope. The blazar is included in a sample of active galactic nuclei (AGN) observed with the POLAMI program \citep{Agudo2014}, which provides both the flux density and polarization measurements at 3~mm and 1.3~mm started in 2009. According to the data during the perion 2007-2021 the degree of linear polarization at 1.3~mm changes from 1\% to 15\%, with circular polarization detections at a level of 1-2\% (private communication).      

The quasar is known as a~violently variable object at optical wavelengths. According to the SMARTS optical/IR monitoring program of blazars \citep{Bonning2012}, \nrao has a fractional variability amplitude of $\sim$68\% and $\sim$35\% at optical and IR wavelengths, respectively, with an average magnitude of $\sim$14 at K band ($\lambda_{\rm eff}\sim$2.2~$\mu$m) and $\sim$17.5 in R band ($\lambda_{\rm eff}\sim$658~nm). \cite{Foschini2006} have reported a short hard X-ray flare detected serendipitously by the IBIS/ISGRI detector on board INTEGRAL on 2004 February 17, although the timescale of the flare, $<$1~hr, questions the association with the FSRQ. However, a moderate increase of the linear polarization at 2\,cm contemporaneous with the flare was observed as well. The recent 10-yr Fermi LAT catalog, 4FGL \citep{4FGL}, lists the average $\gamma$-ray flux of the source at 0.1-300~GeV as (1.8$\pm$1.2)$\times$10$^{-7}$ phot~cm$^{-2}$s$^{-1}$. During this period \nrao
underwent the highest amplitude $\gamma$-ray outburst in 2010 October--November, when the $\gamma$-ray flux rose to 10$^{-6}$ phot~cm$^{-2}$s$^{-1}$ \citep{Karen2014}. This $\gamma$-ray activity coincided with a significant brightening at optical wavelengths and an increase of the optical linear degree of polarization up to 15\%\footnote{ \url{http://www.bu.edu/blazars/VLBA\_GLAST/1730.html}}. 

    \begin{figure*}[t!]
    \centering
    \includegraphics[width=0.99\linewidth]{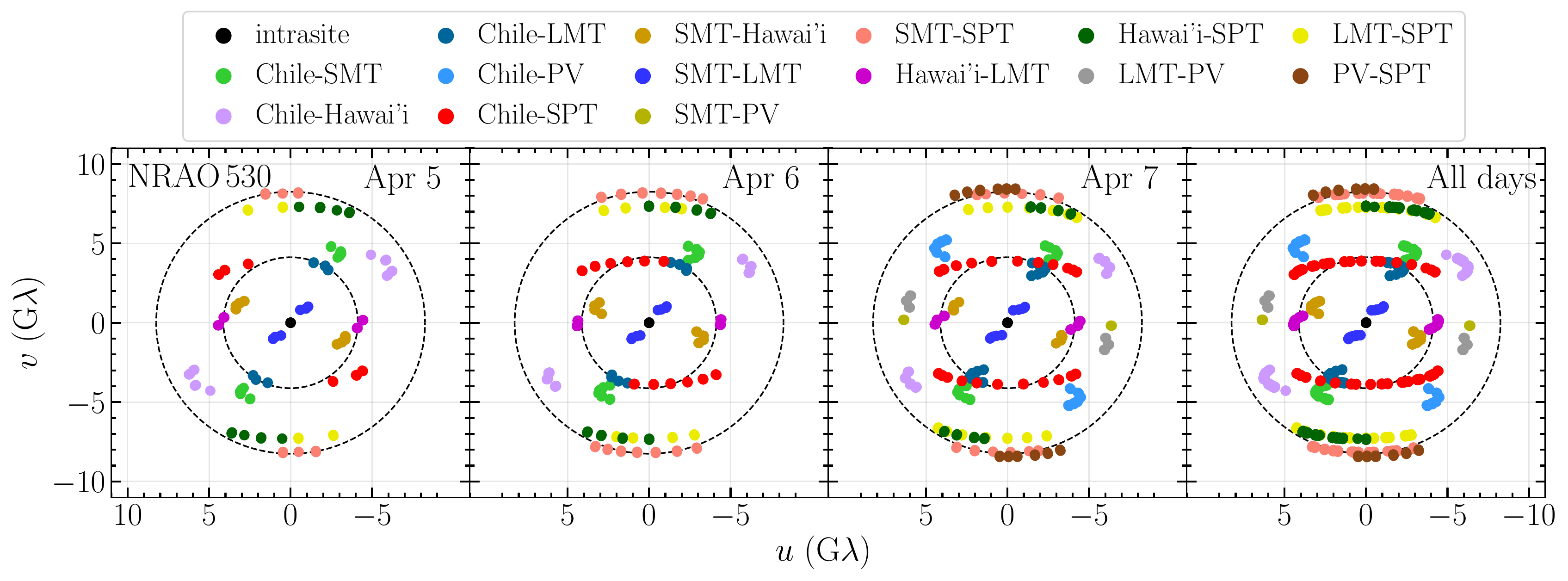}
    \caption{EHT ($u,v$)-coverage of the \nrao observations on 2017 Apr 5, 6, 7, and all days aggregated. Each colored point corresponds to a single VLBI scan of 3-4 min. ALMA participated in the observations on 2017 Apr 6, 7. Dashed circles indicate the fringe spacing of 50\,$\mu$as and 25\,$\mu$as. "Chile" represents the stations ALMA and APEX. "Hawai'i" represents the stations SMA and JCMT.}
    \label{fig:coverage}
\end{figure*}

\nrao possesses a highly relativistic jet. The source is intensively monitored with the Very Long Baseline Array (VLBA) at 15~GHz within the MOJAVE program \citep[e.g., ][]{Lister2016} and at 43~GHz within the VLBA-BU-BLAZAR program \citep{Jorstad2017, Weaver2022}. VLBA images show a radio jet dominated by the core located at the southern end of the jet. At 15~GHz the jet extends up to 10~mas to the north, with a weak feature observed $\sim$25~mas from the core. Based on 11 moving features, \cite{Lister2019} report a median apparent speed of 12.3$\pm0.6~c$, with a maximum apparent speed of 27.3$\pm1.0~c$. At 43~GHz the jet contains a stationary feature located $\sim$0.3~mas from the core and curves to the northwest $\sim$0.8~mas from the core. 8 moving knots detected at 43~GHz over 2007-2018 exhibit a wide range of apparent speeds, from 2.6$\pm0.5~c$ to 37.2$\pm0.2~c$ \citep{Weaver2022}, with 3 knots revealing a strong acceleration after passing the stationary feature, while knot B3 ejected in the beginning of 2010 demonstrates both  accelerations from 4.1~c to 7.7 c and to 33.7~c and then a deceleration to 8.5~c. Using properties of the light curves of moving knots and their apparent speeds, \cite{Jorstad2017} have estimated the average Lorentz and Doppler factors of the jet to be $\Gamma\sim$8 and $\delta\sim$9, respectively, with a jet viewing angle of $\sim$3$^\circ$. These values of the Lorentz and Doppler factors are close to the minimum values obtained by \cite{Liodakis2018} for the jet of \nrao. These authors estimated the variability of the Doppler factor
by analyzing the 15~GHz light curve provided by the Owens Valley 40\,m Radio Telescope after taking into consideration brightness temperature arguments and the apparent speeds of knots at 15~GHz; this resulted in $\Gamma_{\rm max}\sim$40 and $\delta_{\rm max}\sim$21.

On kiloparsec scales the quasar shows a weak two-sided jet elongated in the east-west direction over $2.5''$, with a very diffuse eastern part and a bright, compact  
knot on the western side, located about $1''$ from the core \citep{Kharb2010}. The kiloparsec-scale jet is almost perpendicular to the parsec scale jet and has a complex radio morphology: the eastern part is similar to that expected for Fanaroff-Riley I (FRI) type radio galaxies, while the western part can be classified as FRII type. 

\nrao has a redshift of $z=0.902$ \citep{Junkkarinen1984}, for which 100\,$\mu$as corresponds to a linear distance of 0.803 pc \citep[$H_0 = 67.7$ km s$^{-1}$ Mpc$^{-1}$, $\Omega_m = 0.307$, and $\Omega_\Lambda = 0.693$]{Planck2016}. The source contains a supermassive black hole (SMBH), the mass of which is currently uncertain, with estimates ranging from  3$\times$10$^8 M_\sun$ \citep{MKeck2019} 
to 2$\times$10$^9 M_\sun$ \citep{LianLiu2003}.

Due to its brightness, compactness, and close location on the sky to the Galactic Center,
\nrao is frequently used as a calibrator in very-long-baseline interferometry (VLBI) observations of the radio source Sagittarius A$^*$ (\sgra) located in the center of the Milky Way \citep[e.g., ][]{Rusen2011}. In this paper we report on the VLBI observations of \nrao in 2017 April, when the source was employed as a calibrator during the Event Horizon Telescope (EHT) observation of \sgra at 1.3~mm \citep{SgraP1}. During the EHT observation \nrao was in a moderate activity state at $\gamma$-ray energies with the flux $\sim$1$\times$10$^{-7}$phot~cm$^{-2}$s$^{-1}$ as well as at optical (R band magnitude $\sim$17.8)\footnote{\url{http://www.bu.edu/blazars/VLBA\_GLAST/1730.html}} and radio wavelengths (the flux density at 37~GHz $\sim$3.7~Jy)\footnote{\url{http://www.metsahovi.fi/AGN/data/}}. According to the VLBA monitorings at 43~GHz \citep{Weaver2022} and 15~GHz \citep{Lister2019} no superluminal components were ejected in 2017. The EHT observation of \nrao provides unique insight into the sub-parsec scale structure of the quasar with micro-arcsecond resolution, which are important for understanding the blazar physics in general and properties of the \nrao jet in particular. The outline of the paper is as follows: in Section \ref{sec:observations} we describe the EHT observations of \nrao, their reduction and general data set properties, in Sections \ref{sec:imaging} and \ref{sec:polar} we present results of total intensity and linear polarization (LP) imaging, respectively, in Section \ref{sec:discussion} we discuss our results, and in Section \ref{sec:conclusions} we summarize our findings.
    
\section{Observations and Data Calibration}
\label{sec:observations}

\begin{figure}[t!]
    \centering
    \includegraphics[width=0.37\paperwidth]{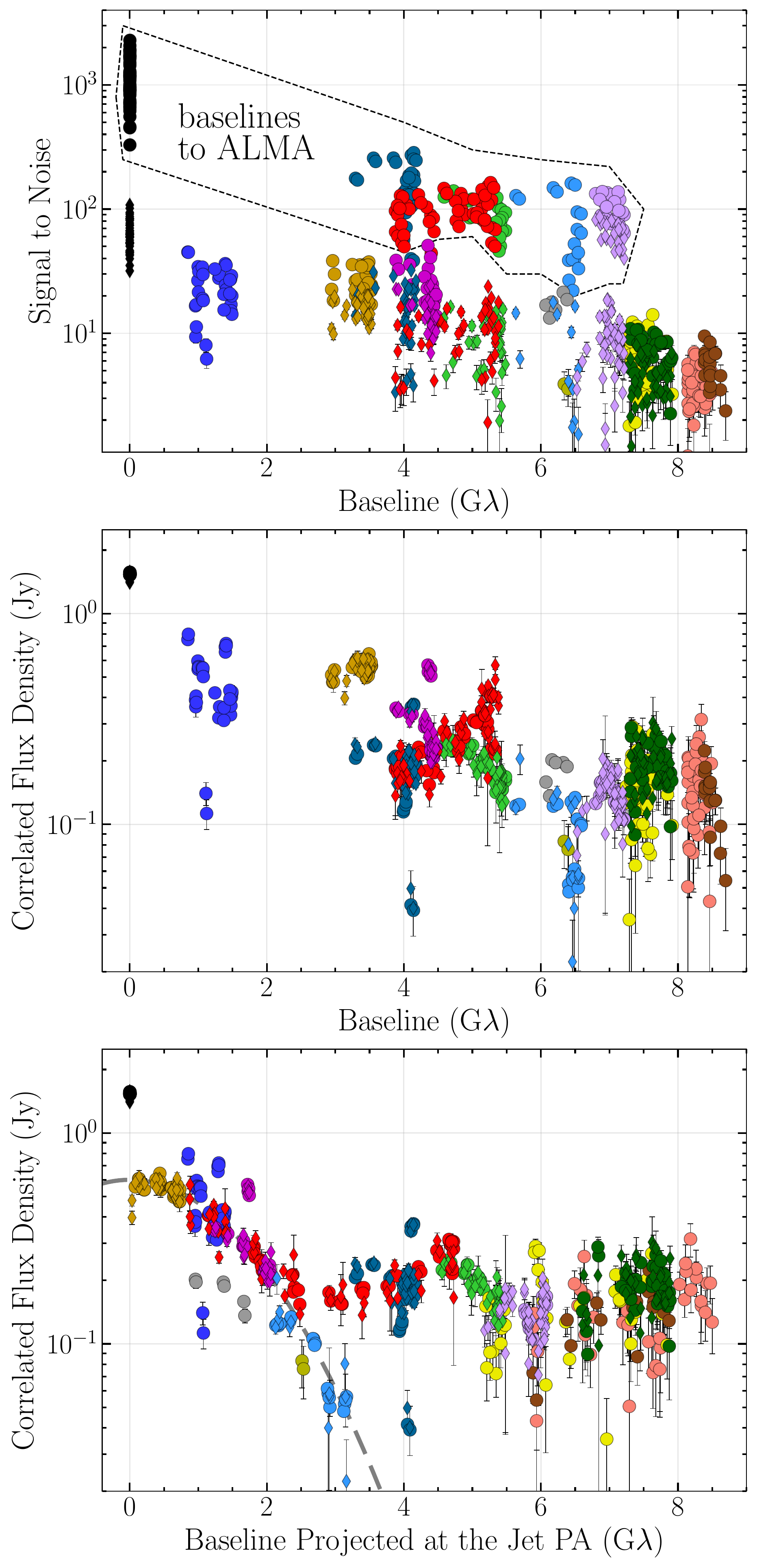}
    \caption{\textit{Top:} S/N of \nrao detections. Colors follow the convention of Figure \ref{fig:coverage}. The entire data set (2017 Apr 5-7, both frequency bands) is shown. The data were averaged coherently in 240\,s intervals. The circles denote primary baselines, while the diamonds denote redundant baselines (i.e., obtained with APEX or JCMT, rather than with ALMA or SMA). The cluster of S/N $> 100$ circles corresponds to ALMA baselines.
    \textit{Middle:} Same as the top panel, but for the correlated flux densities (visibility amplitudes) in the a~priori calibrated data set. \textit{Bottom:} Visibility amplitudes as a function of baseline length projected at the jet position angle (PA) of -25 deg (see Section \ref{sec:imaging}). Dashed line represents a Gaussian component of full width at half maximum (FWHM) 55\,$\mu$as and 0.6\,Jy amplitude. }
    \label{fig:amplitudes}
\end{figure}

The EHT observed \nrao on three consecutive nights on April 5-7 during the 2017 campaign (MJD 57848-57850), as a calibrator of Sgr A$^*$ \citep{SgraP1,SgraP2}. The observations were performed with the full EHT 2017 array of eight telescopes located at six geographical sites: the Atacama Large
Millimeter/submillimeter Array (ALMA)\footnote{ALMA joined on Apr 6 and 7, operating as a phased array of on average 37 dishes of 12\,m diameter, in a compact configuration with longest baselines not exceeding 300\,m length \citep{Goddi2019,wielgus22}.} and the Atacama Pathfinder Experiment (APEX) telescope in Chile; the Large Millimeter Telescope Alfonso Serrano (LMT) in Mexico; the IRAM 30\,m telescope (PV, joined only on Apr 7) in Spain; the Submillimeter Telescope (SMT) in Arizona; the James Clerk Maxwell Telescope (JCMT) and the Submillimeter Array (SMA) in Hawai'i; and the South Pole Telescope (SPT) in Antarctica. Two 2\,GHz wide frequency bands, centered at 227.1\,GHz (LO), and 229.1\,GHz (HI), were recorded. The observations were carried out using dual feeds, right-hand and left-hand circularly polarized (RCP ad LCP, respectively), for all stations other than ALMA and JCMT. ALMA recorded dual linear polarization, which was subsequently converted at the correlation stage to a circular basis by {\tt PolConvert} \citep{Marti_2016,Goddi2019}. The JCMT only observed the RCP component, which we utilized to approximate total intensity (this approximation is correct as long as the fractional circular polarization of the source can be neglected), and we typically omit JCMT baselines for the polarimetric imaging. The EHT array setup is detailed in \citet{EHT2019p1,EHT2019p2}.

Recorded signals were correlated at the MIT Haystack Observatory and the Max-Planck-Institut f\"{u}r Radioastronomie, Bonn. Subsequent data reduction procedures are described in \citet{EHT2019p3,Blackburn_2019,Janssen2019}. There were minor updates to the calibration with respect to the EHT results published earlier \citep{EHT2019p1,Kim2020}, particularly regarding the telescope sensitivity estimates and complex polarimetric gains calibration. These updates are identical to those described in the EHT \sgra\ publications \citep{SgraP1,SgraP2}. Flux density on the short intra-site baselines (ALMA-APEX and SMA-JCMT) was gain-calibrated to the simultaneous ALMA-only flux density of 1.6\,Jy, reported by \citet{Goddi2021}, providing an additional constraint on the amplitude gains for the stations with a co-located partner \citep[network calibration;][]{Blackburn_2019}.
The polarimetric leakage calibration follows procedures outlined in \citet{PaperVII}, where D-terms of stations with an intra-site VLBI baseline (ALMA, APEX, SMA, JCMT) were calculated through a multi-source fitting procedure \citep{MartiVidal2021}, and several estimates for D-terms of remaining stations were reported. Here we employ the fiducial D-terms given by \citetalias{Issaoun2022} based on the analysis presented in \citet{PaperVII}. Some imaging algorithms that we use ignore the leakage calibration and estimate D-terms independently, providing an additional consistency check of the calibration procedures and their impact on the resulting images (see Section \ref{sec:imaging}).

\begin{figure*}[t!]
    \centering
    \includegraphics[width=0.99\linewidth]{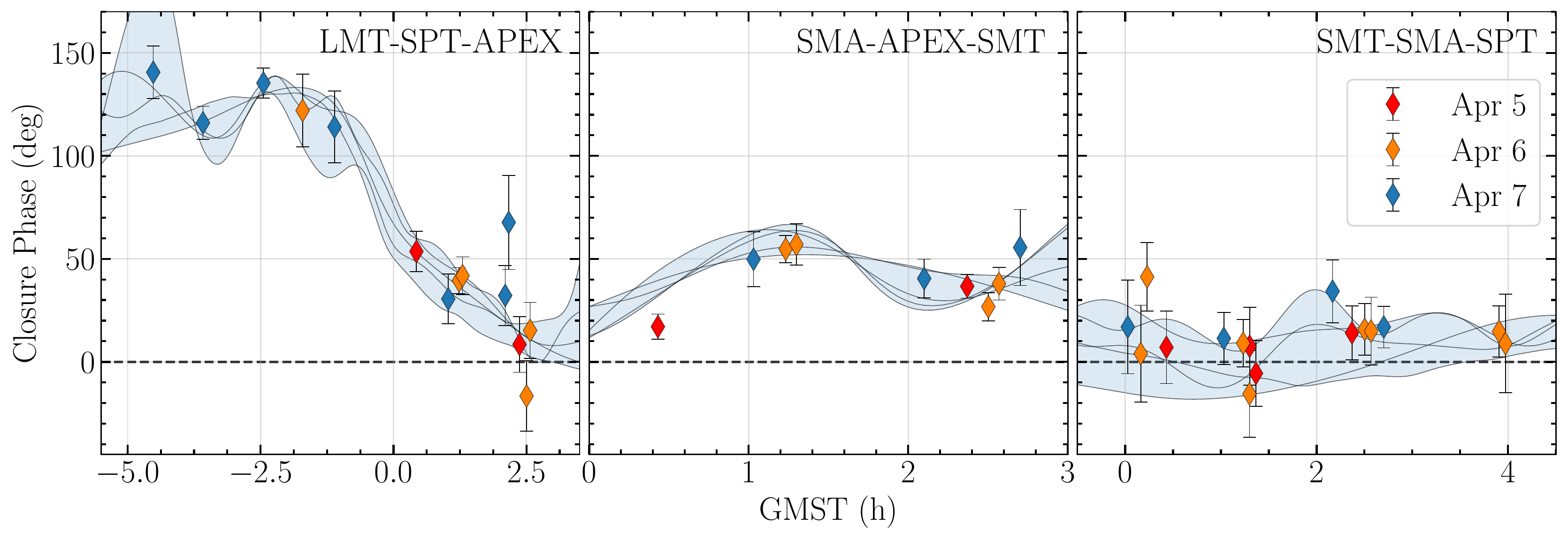}
    \caption{\nrao closure phases measured on a selected non-trivial triangles. LO band data points averaged in 240\,s intervals are shown. There is no statistically significant indication of structural variation between the 3 consecutive observing days. Predictions of the models obtained with a variety of imaging algorithms, discussed in Section~\ref{sec:imaging}, are indicated with thin black lines. The blue-shaded region indicates the range of predictions from different reconstructions presented in this paper. }
    \label{fig:cphases}
\end{figure*}

The $(u,v)$-coverage of the EHT observations of \nrao is given in Figure~\ref{fig:coverage}. We show the coverage on each individual day, as well as aggregated over the entire observing campaign. The longest projected baseline reaches 8.66\,G$\lambda$ for SPT-PV, which determines the instrumental angular resolution of 24~${\rm\mu}$as, measured as the minimum fringe spacing. Inspection of the data revealed no indication of the source structure evolution during the observations spanning only about 52\,h on 2017 Apr 5$-$7, with the total on-source integration time of 76 minutes (see Figures \ref{fig:amplitudes}-\ref{fig:cphases}). For that reason, the entire data set was used simultaneously for static imaging and model fitting.

The \nrao data set self-consistency has been rigorously verified as a validation of the EHT Sgr\,A$^\ast$ results, presented in \citet{SgraP2}. The signal-to-noise ratio (S/N) of individual \nrao detections (band-averaged in frequency, scan-averaged in time) typically exceeds 100 on baselines to ALMA, and is about an order of magnitude lower on remaining baselines; see the top panel of Figure~\ref{fig:amplitudes}. We estimated additional 2\% of complex visibility uncertainties due to variety of non-closing effects (e.g., residual polarimetric leakage, frequency and time smearing), see \citet{EHT2019p3,SgraP2}. Inspecting the calibrated visibility amplitudes in the middle panel of Figure~\ref{fig:amplitudes}, we notice that the shortest inter-site baseline (LMT-SMT) provides a lower limit on the compact emission flux density corresponding to about 0.6\,Jy, which implies the presence of up to 1\,Jy of mas scale flux that is resolved out by the EHT but measured on the intra-site VLBI baselines and by the ALMA-only observations. Additionally, in the bottom panel of Figure~\ref{fig:amplitudes} we present the measured visibility amplitudes as a function of the baseline length, projected at the jet PA (see Section \ref{sec:imaging}). We notice that the amplitudes on baselines shorter than about 2.5\,G$\lambda$ can be approximated by a Gaussian of FWHM 55\,$\mu$as and 0.6\,Jy amplitude, roughly informing us about the size and the total flux density of the observed compact feature. Measurements on baselines longer than 2.5\,G$\lambda$ indicate a deviation from the Gaussian model, in agreement with the resolved source consisting of more than one component.
Both EHT \sgra calibrators, \nrao and J1924-2914 \citepalias{Issaoun2022} are fairly compact and bright sources at 230\,GHz. During the week-long EHT observing run, their source structure and flux density did not show significant variations (see Figure \ref{fig:cphases}). These properties allowed us to use both sources for quantifying the effects of the residual errors in the characterization of the antenna gains. Mitigating such effects was crucial for discerning between the rapid intrinsic variability of \sgra and the apparent flux density variations produced by imperfect calibration \citep{SgraP3,SgraP4,wielgus22}. Details about the procedures and methods used to obtain the antenna gains from the two calibrators and the subsequent transfer to the \sgra data sets are reported in Section 3.2 in \citet{SgraP2}.

\section{Total Intensity Imaging}
\label{sec:imaging}

For the imaging of \nrao we have utilized data calibrated by the \texttt{EHT-HOPS} pipeline \citep{Blackburn_2019}, combined over all days (2017 Apr 5-7) and both frequency bands (227.1 and 229.1 GHz). Non-zero closure phases shown in Figure \ref{fig:cphases} reveal resolved and non-trivial structure of the source. We used three analysis methods across five different implementations. These are: imaging via inverse modeling (the traditional CLEAN method), forward regularized maximum likelihood (RML) modeling (\texttt{eht-imaging} and {\tt SMILI} algorithms), which are described in detail in \cite{EHT2019p4}, as well as two posterior exploration methods based on a Markov chain Monte Carlo (MCMC) scheme -- a~\mbox{D-term} Modeling Code \citep[{\tt DMC};][]{Dom2021} and a dedicated Bayesian image reconstruction framework {\tt Themis} \citep{Broderick2020}. For each of the method the freedom to decide resolution, field of view, or amount of data coherent averaging was left to the expert imaging sub-teams, who made such decisions based on the specific properties of each algorithm. As an example, MCMC methods tend to average data more to reduce the computational complexity of the likelihood evaluation. Below we give a short description of each method, with some details on the particular application to the \nrao imaging.

\subsection{{\tt DIFMAP}}
For CLEAN imaging we have used the {\tt DIFMAP} software package \citep{Difmap}. We have constructed multiple manual images prior to the final process to obtain a set of CLEAN windows corresponding to the source structure and agreeing best with the data. This window set includes a large circular Gaussian with a radius of 1~mas, located in the center of the map, which models the large-scale extended emission present in the source. The standard process of multiple consecutive CLEAN iterations, along with self-calibration procedures, was employed to achieve the best normalized $\chi^2$ values between the data and the model for the closure phases, closure amplitudes, and visibilities. We have used data averaged in 10\,s segments, a field of view of 1024$\times$1024 pixels with a pixel size of 2~$\mu$as and uniform weighting to optimize the image resolution.

\subsection{{\tt eht-imaging}}
\label{sec:ehtim}
The \texttt{eht-imaging} code \citep{Chael_2016, Chael_2018} implements the RML-based approach to imaging. In this method, an updated image is being proposed at every iteration, while the algorithm optimizes a cost function comprised of the error term describing consistency with the data (predominantly through closure phases and log closure amplitudes; \citealt{Blackburn2020}), and regularization terms, similar to Bayesian priors. The latter may promote image properties such as smoothness, sparsity, compactness, or consistency with the imaging prior -- here chosen to be a 60\,$\mu$as FWHM circular Gaussian. For the \nrao analysis we assumed a field of view of 200\,$\mu$as and a 64$\times$64 pixel grid, and used scan-averaged data. The algorithm proceeds with an iterative sequence of gain calibration and cost function optimization.

\subsection{{\tt SMILI}}
{\tt SMILI} \citep{Akiyama_2017a} is another RML imaging library that reconstructs interferometric images utilizing a similar set of regularizers including weighted-$\ell_1$ ($\ell_1^w$), total variation (TV), total squared variation (TSV), and the maximum entropy regularizer (MEM). Prior to the imaging, we rescale the intra-site baseline flux density of \nrao to 0.9\,Jy to remove the contributions from the large-scale extended emission. During the imaging we adopt a field of view of 300\,$\mu$as and a 150$\times$150 pixel grid, initializing the algorithm with a circular Gaussian prior, and using 10\,s averaged data set. Subsequently, the imaging algorithm minimizes the regularized cost function. After 1000 iterations, self-calibration of the stations gains is performed and the image prior is updated. This imaging and self-calibration loop is repeated until the reconstruction converges.

\subsection{{\tt DMC}}

{\tt DMC} formulates the imaging problem in terms of posterior exploration, which is carried out using a Hamiltonian Monte Carlo sampler as implemented in the PyMC3 Python package \citep{Salvatier_2016}. The output of {\tt DMC} is a collection of samples from the joint posterior over the image structure and calibration quantities. Rather than carrying out an iterative self-calibration procedure, {\tt DMC} fits the complex gains at every station simultaneously with the full Stokes image structure. For \nrao imaging, a field of view of 300\,$\mu$as and a~30$\times$30 pixel grid were used, the data were scan-averaged, and D-terms were fixed following \citetalias{Issaoun2022}. A~comprehensive description of the {\tt DMC} model and implementation can be found in \cite{Dom2021}.

\subsection{ {\tt Themis} }
\label{sec:themis}
{\tt Themis} is a framework for fitting VLBI data with a~Markov chain Monte Carlo (MCMC) posterior space exploration method \citep{Broderick2020}. For \nrao data analysis the image was modeled with a~spline raster defined using 6$\times$5 control points, and an adaptively selected field of view \citep{Broderick2020b}. The data were scan-averaged. We solved for D-terms with {\tt Themis} in order to provide a consistency test for the fiducial D-terms of \citet{PaperVII} and \citetalias{Issaoun2022}, see Section \ref{app:dterms}.

\subsection{Fiducial Images of NRAO\,530 }
\label{sec:fiducial_images_I}
\begin{figure*}[t!]
    \centering
    \includegraphics[width=0.2\textwidth]{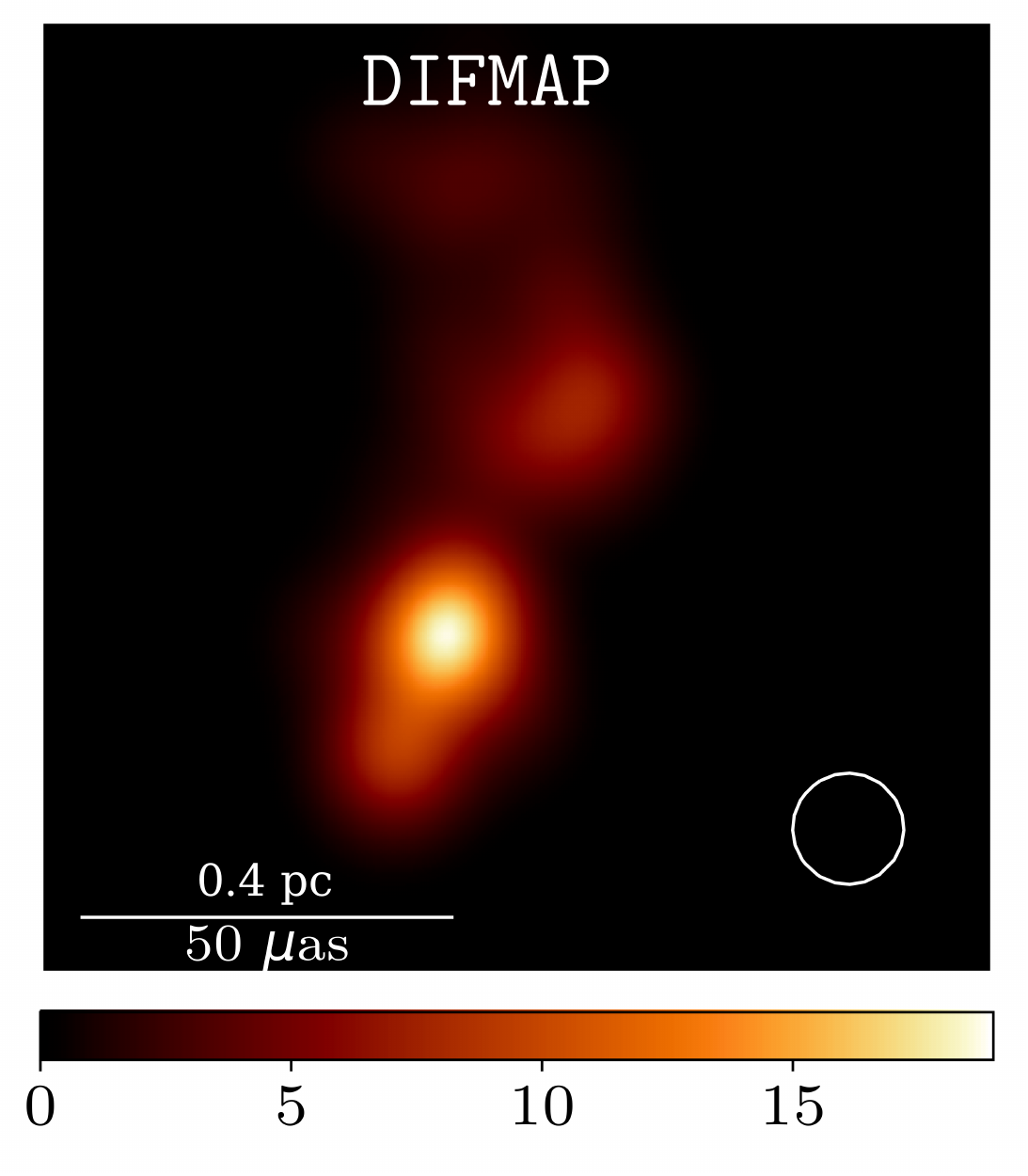}\hspace{-0.2cm} 
    \includegraphics[width=0.2\textwidth]{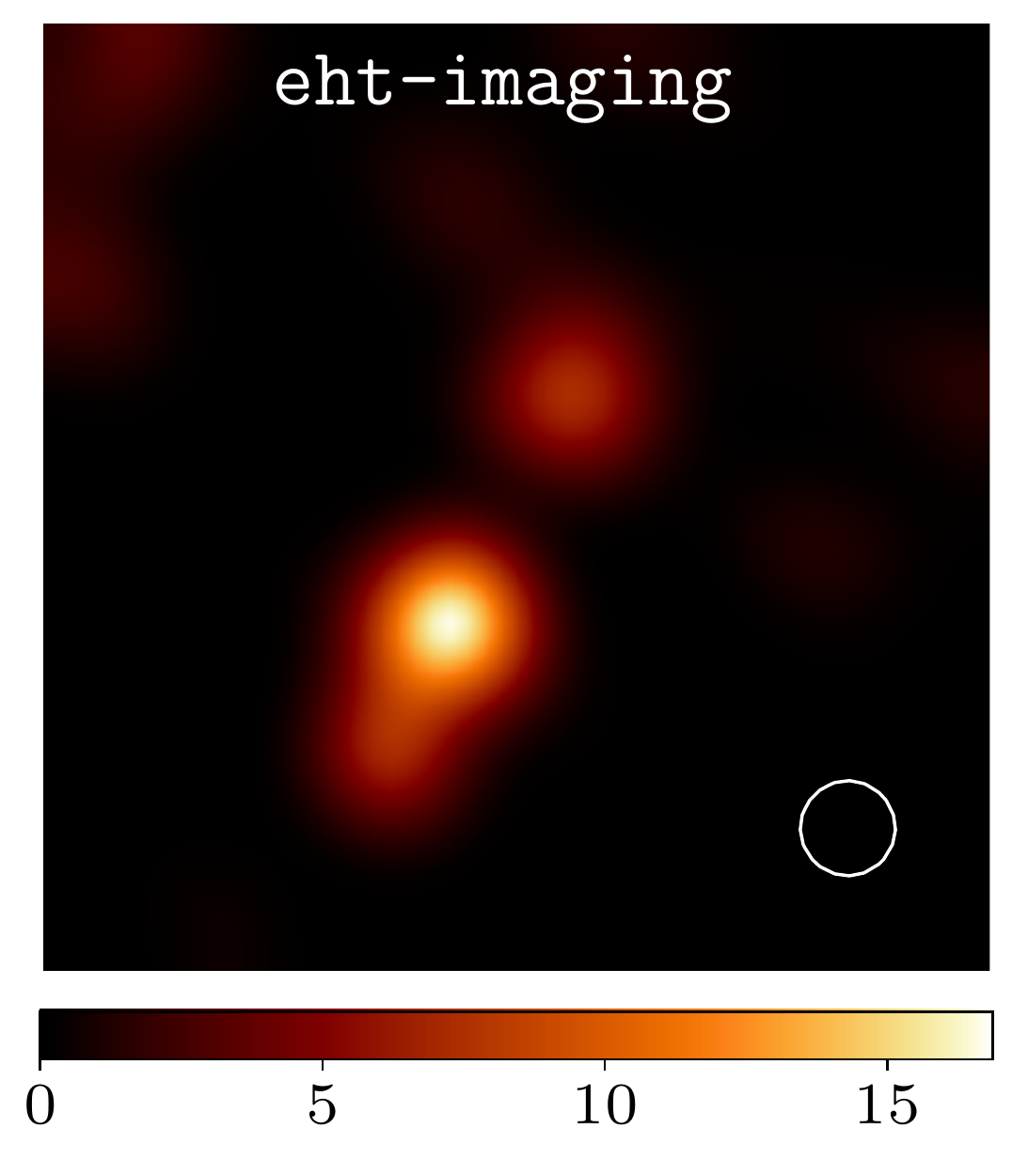} \hspace{-0.25cm}   
    \includegraphics[width=0.2\textwidth]{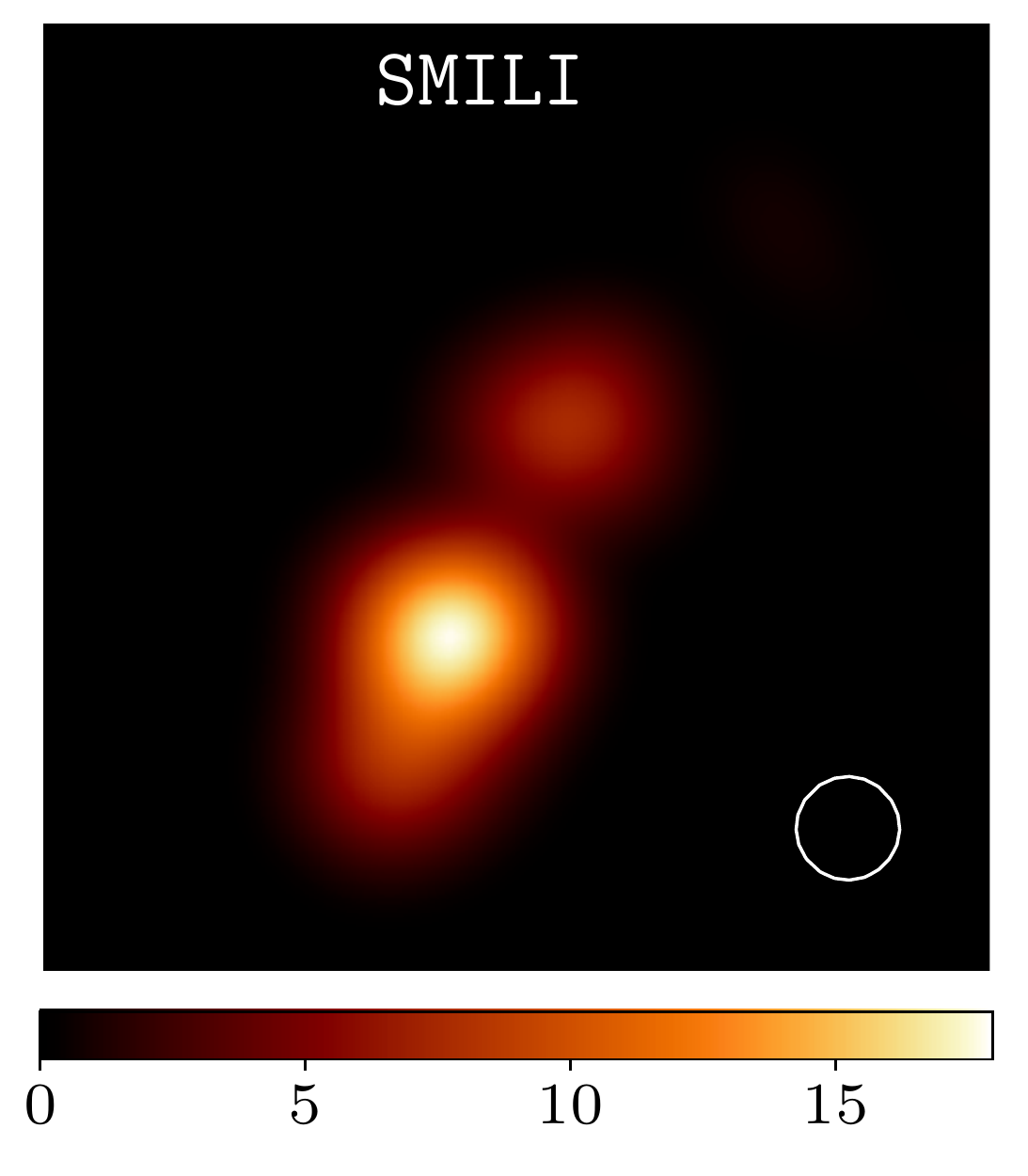}\hspace{-0.1cm}   
    \includegraphics[width=0.2\textwidth]{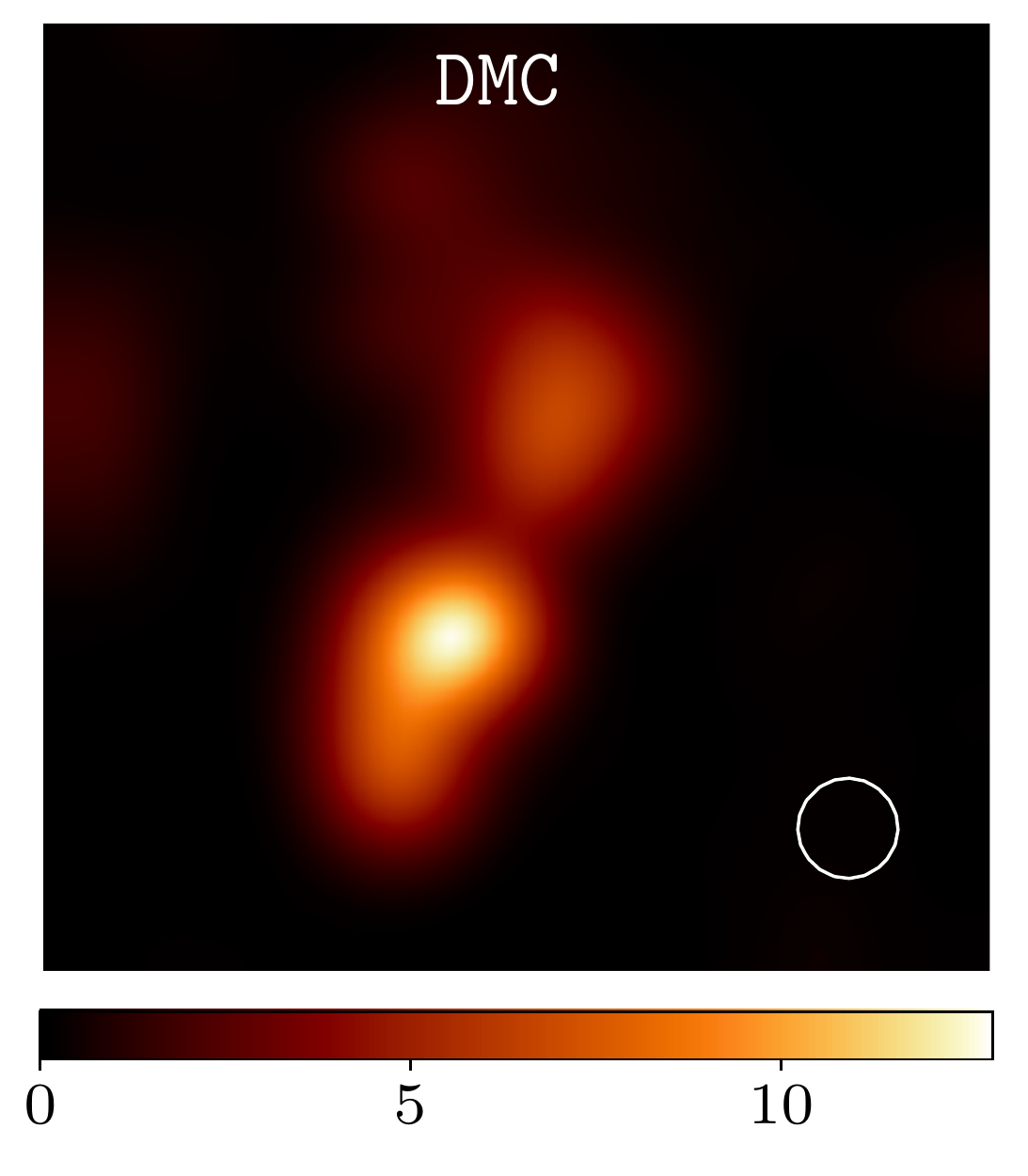} \hspace{-0.2cm}   
    \includegraphics[width=0.2\textwidth]{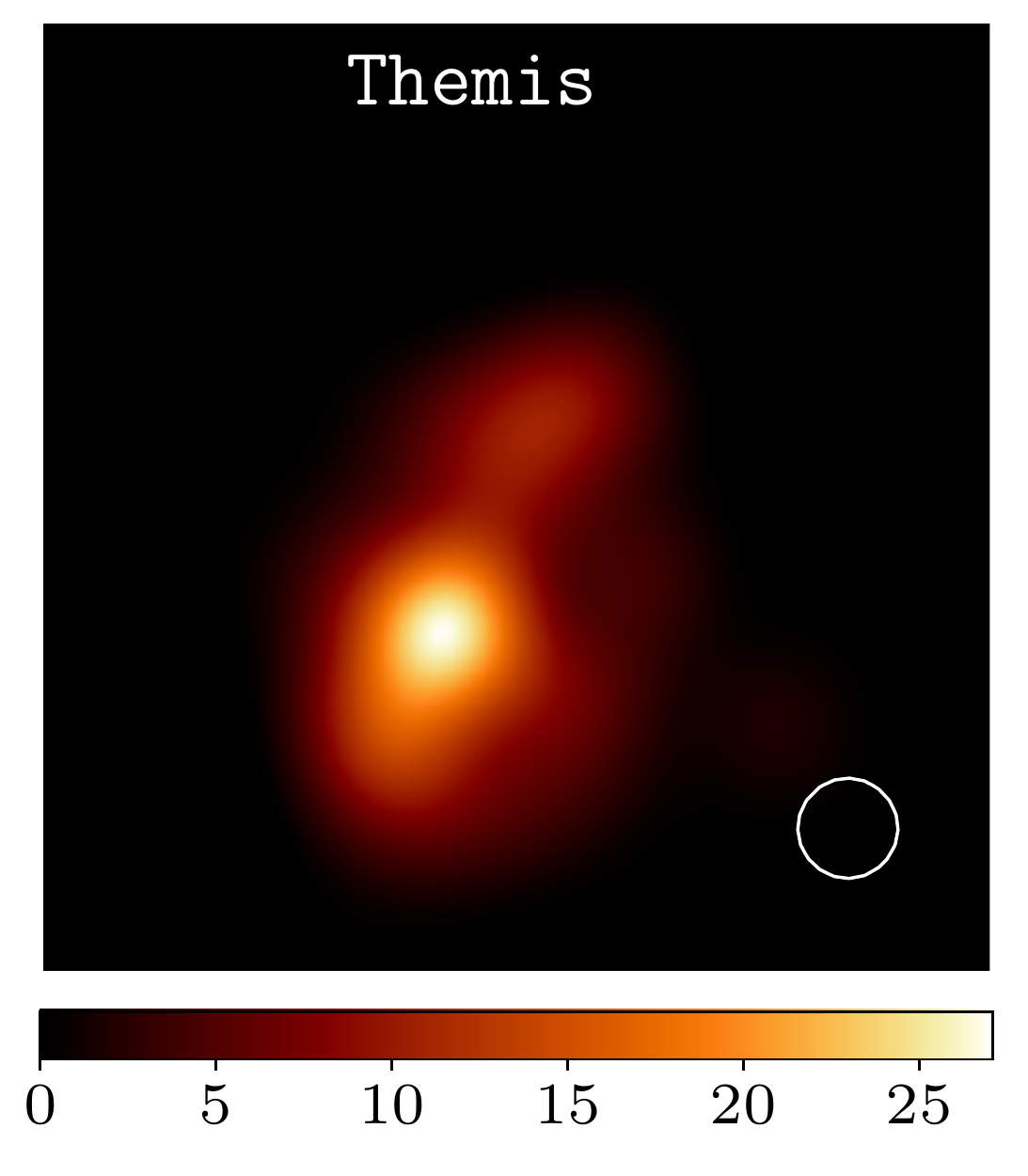}   
    \vspace{-0.5eM}\\
    {\large Brightness Temperature ($10^9$ K)}
    \caption{Images of \nrao from the 2017 April EHT observations produced using {\tt DIFMAP}, {\tt eht-imaging}, {\tt SMILI}, {\tt DMC} and {\tt Themis}, shown within the adopted field of view of 128\,$\mu$as. To simplify visual comparisons, the models were blurred to similar effective resolutions $\sim 15\,\mu$as, indicated in the bottom right corner of each image.  
    \vspace{0.5eM}\\
}
    \label{fig:images}
\end{figure*}

Figure~\ref{fig:images} presents final (fiducial) total intensity images obtained by the methods discussed in Sections \ref{sec:ehtim}-\ref{sec:themis}. For the presentation purposes all images are shown using the same field of view of $128 \times 128\,\mu$as, more narrow than the field of view typically used for the imaging. The contribution of emission outside of this region is relatively small and consistent with the image noise level, which in case of the EHT is rather high and dominated by the systematic uncertainties related to sparse $(u,v)$-coverage. There is a high degree of agreement between the different imaging methods in terms of the source structure and the relative brightness of individual features. All images show the jet to be elongated from the southeast to the northwest over $\sim$60~$\mu$as. A visual inspection suggests that the jet consists of two dominant features (see Figure \ref{fig:1mmImage}), with the brightest one, C0, located at the southeast end of the jet and itself exhibiting substructures C0a and C0b. However, there is also a~difference between the models -- {\tt DIFMAP} and {\tt DMC} images contain a very low-brightness extended feature that turns to the northeast, almost by $\sim$90$^\circ$ to the main direction of the jet, the {\tt eht-imaging} image shows a hint of such a feature, while the {\tt SMILI} and {\tt Themis} methods do not reveal its presence. This discrepancy is likely related to the latter two algorithms promoting sparsity and compactness in the reconstructed images. We have designated this ambiguous feature as C2 in Figure \ref{fig:1mmImage}, marked by a dashed line ellipse. Although the knot appears in the fiducial method-averaged image if we increase the dynamic range, its brightness peak is on the level of the image noise. In addition, {\tt DIFMAP}, {\tt eht-imaging}, and {\tt SMILI} images correspond to very similar peak brightness values, but the {\tt Themis} and {\tt DMC} images indicate somewhat brighter and weaker brightness peaks, respectively. 

Figure \ref{fig:cphases} shows consistency between closure phase measurements on selected triangles and corresponding predictions of the fitted models. Table \ref{tab:chi-squares} provides goodness of fit normalized $\chi^2$ statistics computed for the models shown in Figure \ref{fig:images} for self-calibrated visibility amplitudes ($\chi^2_{\rm AMP}$), closure phases ($\chi^2_{\rm CP}$), and log closure amplitudes ($\chi^2_{\rm logCA}$). The estimated total flux density of the compact source model is also provided. This quantity is particularly challenging to constrain with the EHT array, which in 2017 lacked baselines between 5\,M$\lambda$ and 500\,M$\lambda$ projected length, sensitive to mas-scale emission. Based on the presented statistics, there is no strong preference for any particular model. Therefore, we have combined the final images constructed by the different methods, shown in Figure~\ref{fig:images}, to produce a~fiducial method-averaged total intensity image of \nrao, presented in Figure~\ref{fig:1mmImage}. 
The averaging procedure involves aligning images to maximize their cross-correlation, regridding them to a common resolution and field of view, and taking a standard arithmetic mean on a pixel-by-pixel basis. It follows the practice of \citet{EHT2019p4}, \citet{Kim2020}, \citet{SgraP3}, and \citetalias{Issaoun2022}, and it is designed to make the resulting image domain morphology more robust against method-specific systematics.

\subsection{Total Intensity Component Modeling}
\label{sec:modeling_total_int}

\begin{figure}[t!]
    \centering
    \includegraphics[width=0.4\paperwidth]{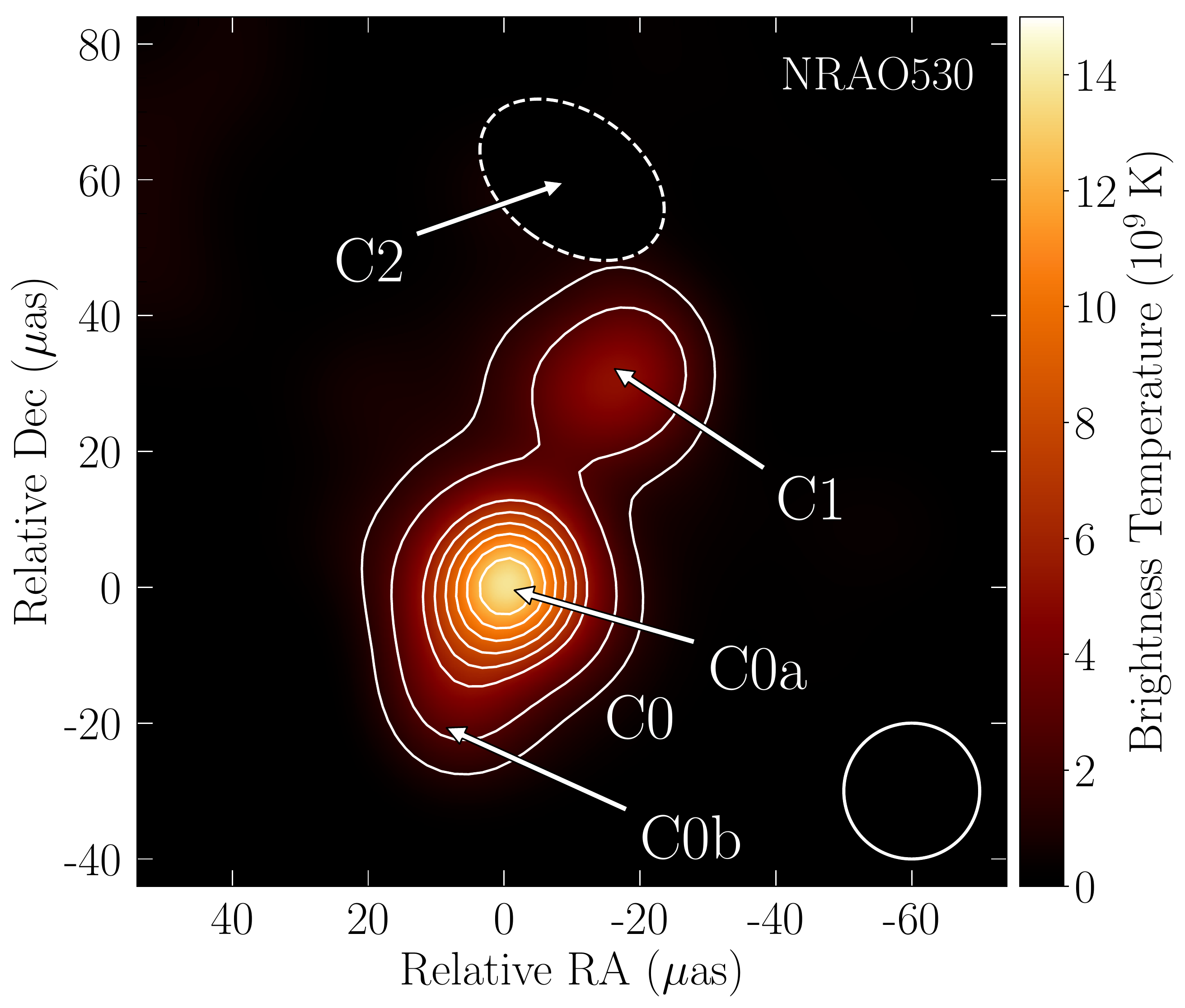}
    \caption{Fiducial EHT total intensity image of \nrao; the color scale corresponds to the brightness temperature; the contours represent the total intensity, with a peak value of $1.4 \times 10^{10}$\,K, starting at 10\% of the peak and increasing in steps of 10\%; the circle in the bottom right corner shows the size of the nominal EHT beam with a FWHM diameter of 20\,$\mu$as. The C2 contour is shown with dashed line as it is below the 10\% level of peak total intensity in the averaged image. }
    \label{fig:1mmImage}
\end{figure}

\begin{table}[h!]
    \centering
    \vspace{5mm}
    \caption{Characterization of the fits to \nrao data.}
    
    \begin{tabularx}{0.41\paperwidth}{cccccc}
    \hline 
    \hline 
         Quantity & {\tt DIFMAP} & {\tt eht-imaging} & {\tt SMILI} & {\tt DMC} & \tt{Themis} \\       
       \hline 
         $\chi^2_{\rm AMP}$ &  0.974 & 0.638 & 0.858 & 0.517 & 1.211 \\
          $\chi^2_{\rm CP}$ &  3.812 & 0.899  & 2.346 & 0.977 & 4.210 \\
          $\chi^2_{\rm  logCA}$ &  1.513 & 1.137 & 2.727 & 0.943 & 2.844 \\
          Flux (Jy) &  1.030 & 0.958 & 0.899 & 1.585 & 1.120 \\
         \hline
         \hline 
    \end{tabularx}
    
    \label{tab:chi-squares}
    NOTE -- 
    For the $\chi^2$ calculation the original thermal error budget was inflated by a non-closing systematic error amounting to 2\% of the complex visibilities.
\end{table}
\begin{table}[h!]
    \centering
    \vspace{5mm}
    \caption{Parameters of total intensity components.}
    \setlength{\tabcolsep}{3pt}
    \begin{tabularx}{0.4\paperwidth}{cccccc}
    \hline 
    \hline 
         Knot &  $S$ (mJy) & $R$ ($\mu$as) & $\Theta$ (deg) & $a$ ($\mu$as) & $T_{\rm b}$ (10$^{10}$\,K) \\
       \hline  
         (1)    &  (2) & (3)  & (4)  & (5)  & (6) \\
       \hline 
       \multicolumn{6}{c}{{\tt DIFMAP} (Gaussian modeling)  }\\
       \hline
         C0a &  278$\pm$29 & 1.8$\pm$4 &85.6$\pm$71 & 16.3$\pm$4.2 & 2.3 \\
         C0b &   61$\pm$15 &21.9$\pm$4 &149.6$\pm$13 & 0.0 & \\
         C1 &  124$\pm$15 &35.1$\pm$5 &-26.9$\pm$9 & 15.0$\pm$3.8 & 1.3 \\
         C2 &  106$\pm$53 &57.5$\pm$20 &-5.2$\pm$24  & 33$\pm$16 & 0.2 \\
         BG &  871 & 1.8 &46.8 & 1000 &\nodata \\
         \hline
        \multicolumn{6}{c}{{\tt eht-imaging} (Gaussian modeling)}\\
         \hline
         C0a &  210 & 0.0 &\nodata & 10.6 & 4.4 \\
         C0b &  74 & 18.8 &154.2 & 5.0 & 6.9 \\
         C1 &   108 & 28.5 &-21.2 & 15.8 & 1.0 \\
         C2 & 158 & 44.3 & -23.8 & 42.8$\times$13.6 & 0.6\\
         BG &  1000 & \nodata & \nodata & \nodata &\nodata \\
         \hline
          \multicolumn{6}{c}{{\tt eht-imaging} (image domain, no blur)}\\
         \hline
         C0a &  228 & 0.0 &\nodata & 13.8$\times$12.3 & 3.1 \\
         C0b &  86 & 18.9 &154.8 & 10.3$\times$11.5 & 1.7 \\
         C1 &   110 & 35.8 &-28.0 & 13.0$\times$12.7 & 1.6 \\
         BG &  978 & \nodata & \nodata & \nodata &\nodata \\
         \hline
         \hline 
         \label{tab:modeling_total_intensity}
    \end{tabularx}

The columns are as follows: (1) - designation of knot; (2) - flux density in mJy; (3) - distance from the core in $\mu$as; (4) - position angle with respect to the core in degree; (5) - size in $\mu$as; (6) - observed brightness temperature in 10$^{10}$K.
  \label{tab:fluxmod}
\end{table}
\begin{figure*}[t]
    \centering
    \includegraphics[width=0.99\linewidth]{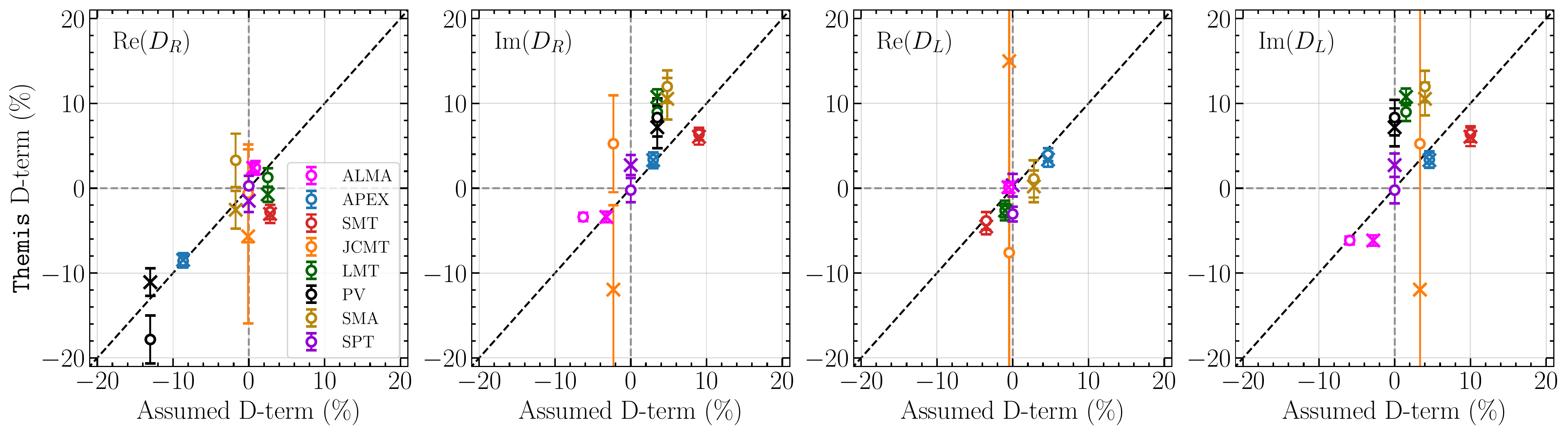}

    \caption{A comparison between the leakage calibration D-terms estimated by \texttt{Themis} from the \nrao data set and D-terms assumed by other imaging methods, following the procedures described in Section \ref{sec:observations}. Empty circles correspond to LO band, crosses correspond to HI band, and error bars represent 1$\sigma$ width of the \texttt{Themis} posteriors.}
    \label{fig:DtermsThemis}
\end{figure*}

We have modeled the visibilities, calibrated to the fiducial image, with circular Gaussian components using the task {\it modelfit} in {\tt DIFMAP}. We also fitted a Gaussian model to the data in the {\tt eht-imaging} framework, using closure-only data products to constrain the resolved source structure. In the latter case the simplest model representing data properties sufficiently well consisted of a single elliptical and 3 circular Gaussian components. Each component (knot) is characterized by a flux density $S$ at distance $R$ from the center of the image with the relative right ascension and declination equal to $(0,0)$, located at the flux density peak of the image, a position angle $\Theta$ with respect to the image center measured from north to east, and a size $a$ corresponding to the FWHM of the Gaussian brightness distribution. Additionally, we performed modeling of the \nrao\ {\tt eht-imaging} results by fitting elliptical Gaussian components to the image domain features. Table \ref{tab:modeling_total_intensity} gives the results of these three separate fitting procedures. To estimate uncertainties of components' parameters we have used an approach employed in high frequency VLBI, e.g., \cite{J05}: the flux density uncertainties correspond to either 10\% of the flux density measurement or the noise level of the map obtained as the result of the {\it modelfit} task in {\tt DIFMAP} depending on whatever is higher, while the position uncertainties are calculated as 1/5-th of the synthetic CLEAN beam corresponding to the ($u,v$)-coverage for compact components, or equal to the beam size for diffuse knots with a size larger than the size of the beam. 

The source consists of two main components, C0 and C1, as indicated in Figure~\ref{fig:1mmImage}. The knots are separated by $\sim$30~$\mu$as, with C0 being a factor of $\sim$3 brighter than C1. We associate C0 with the core of \nrao at 1.3~mm and determine the jet direction on $\mu$as-scales to be $\sim-$28$^\circ$ east of north. All images shown in Figure \ref{fig:images} indicate non-trivial structure in C0, elongated along the jet axis. The models super-resolve the substructure of C0 into components C0a and C0b. In the case of {\tt DIFMAP} the model yields the best fit to C0b as a point source. We interpret the brighter component C0a as the VLBI core. The weaker component C0b is located $\sim$20~$\mu$as southeast from the C0a core, along the same jet axis as the C0a and C1 components. Taking into account a small viewing angle of the jet, it is very unlikely that C0b is a counter-jet component. Instead, we interpret C0b as the result of the opacity stratification in the core, with C0b representing more optically thick part of the jet. The Gaussian component fits show the presence of C2 component, and although uncertainties of the parameters are large, there is a good agreement between the results of Gaussian fits of {\tt DIFMAP} and {\tt eht-imaging} modeling. 

We have calculated the observed brightness temperature of the knots, $T_{\rm B}$, in the same manner as in \citetalias{Issaoun2022} (no cosmological redshift or Doppler corrections accounted for).  According to Table~\ref{tab:fluxmod}, C0 has a higher $T_{\rm B}$ than the C1 component, which supports the interpretation that it is a VLBI core at 1.3~mm. The observed brightness temperature of the core is close to the equipartition value of $T_{\rm eq}\sim 5 \times 10^{10}$\,K \citep{Readhead94}, which implies an intrinsic (plasma frame) brightness temperature  $T'_{\rm B}= T_{\rm B}(1+z)/\delta <<T_{\rm eq}$, if we assume the same Doppler factor of $\delta \sim 9$ on $\mu$as scales as estimated on mas scales \citep{Jorstad2017}. A robust comparison with $T_{\rm eq}$ requires calculation of $T_{\rm B}$ either for an optically thin knot or for the flux density corresponding to the turnover frequency in the spectral energy distribution (SED) of a knot. However, the spectral index obtained for \nrao by \cite{Goddi2021} is steep, $\alpha\sim 0.8$ ($S_\nu\propto\nu^{-\alpha}$), while the source is dominated by the core at 221~GHz, so that even for C0 we can assume that we measure $T_{\rm B}$ either near the turnover frequency or at the optically thin branch of the SED. The brightness temperature is somewhat higher at lower frequencies, where the core is optically thick. This will be discussed in a separate paper devoted to multi-wavelength observations of \nrao (Lisakov et al. in prep.). The results shown in Table~\ref{tab:fluxmod} also contain a large Gaussian component, BG, with a flux density of $\sim$1\,Jy, which characterizes the unconstrained source emission on larger angular scales, overresolved on inter-site EHT baselines.


\section{Linear Polarization Images}
\label{sec:polar}

We have performed linear polarization imaging of \nrao using four out of five algorithms described in Section~\ref{sec:imaging}: {\tt DIFMAP} (CLEAN), \texttt{eht-imaging}, \texttt{DMC}, and \texttt{Themis}. {\tt SMILI} implementation does not currently have the capability to reconstruct polarimetric images. Given the large and poorly constrained systematic uncertainties in the polarimetric calibration of the EHT data at times when ALMA is not participating in observations, only 2017 Apr 6 and 7 data were used, except for {\tt DMC}, which allows for a very flexible calibration of complex polarimetric gains simultaneous with the image reconstruction \citep{Dom2021}, and hence could reliably fit the entire data set. Similarly, except for {\tt DMC} and {\tt Themis}, we disregarded JCMT single polarization data for the LP analysis.   

\subsection{Consistency Test for D-terms}
\label{app:dterms}

For all methods other than {\tt Themis}, we either fixed all polarimetric leakage D-terms following multi-source fitting procedures described in \citet{PaperVII} and \citetalias{Issaoun2022}, or we only solved for the SPT D-terms. The latter resulted in the SPT D-terms being no larger than 1\%, with uncertainties comparable to or larger than the values themselves, which justifies setting SPT D-terms to zero. In the case of {\tt Themis}, we performed full fitting of D-term coefficients simultaneously with the image reconstruction as an additional consistency test. In Figure~\ref{fig:DtermsThemis} we provide a comparison between the assumed and estimated D-terms. The results show a generally high level of consistency, particularly when considering the difficulty of quantifying the systematics present in a fit to a single sparsely sampled data set. JCMT D-terms are constrained very poorly as a consequence of the station operating with a single polarization receiver in 2017. There is a good correlation between the assumed and calculated D-terms, corresponding to a Pearson coefficient of 0.86 for the vector of 64 D-terms (ignoring the uncertainties) and median residual of 2\% on a single complex D-term. Altogether, the consistency test results support fixing D-terms for the model fitting and data analysis.


\begin{figure*}[t]
    \centering
    \includegraphics[height=14cm,trim={-1.5cm 0.2cm 0.2cm 0.0cm},clip]{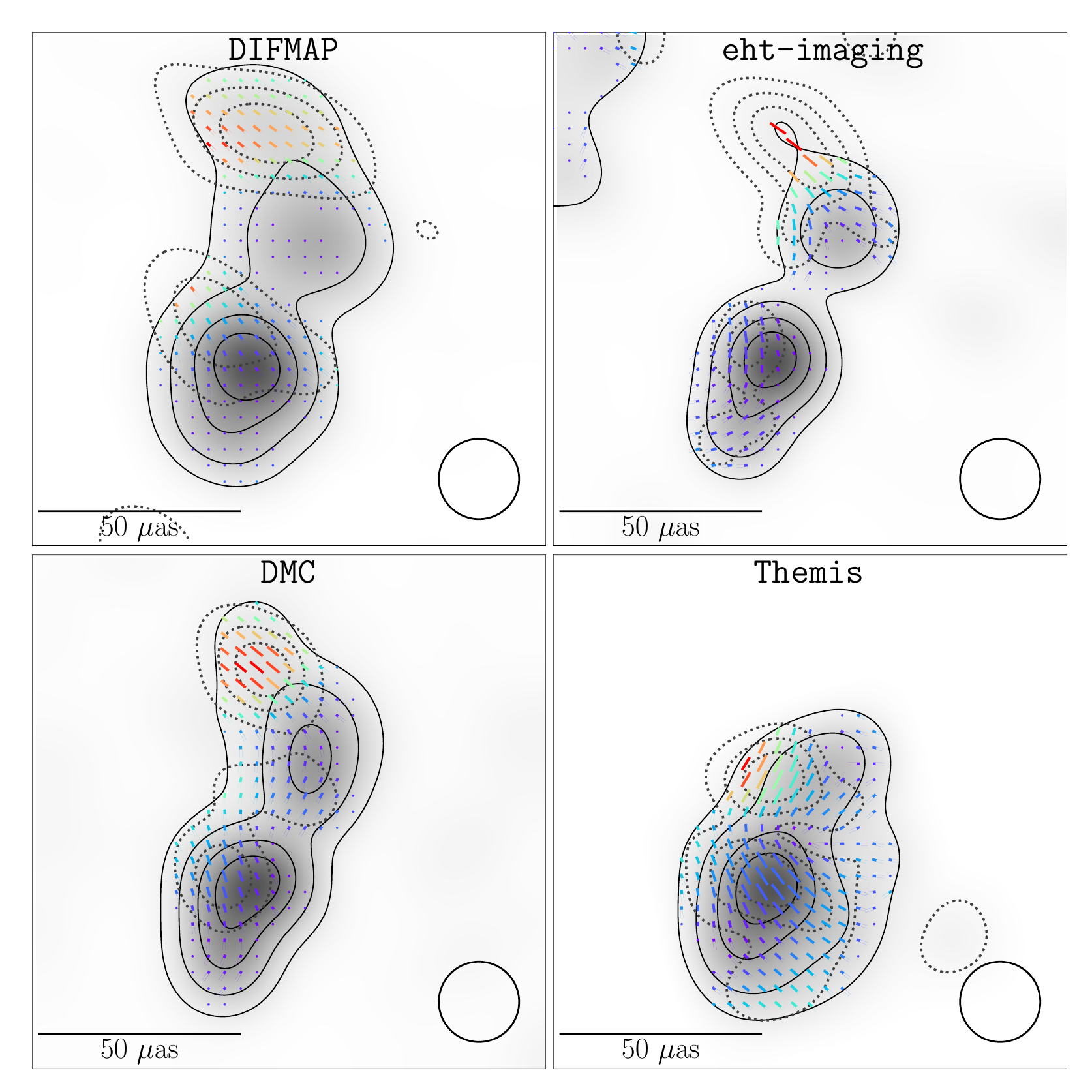}
    \includegraphics[height=14cm,trim={0.2cm 0.2cm 1cm 0cm},clip]{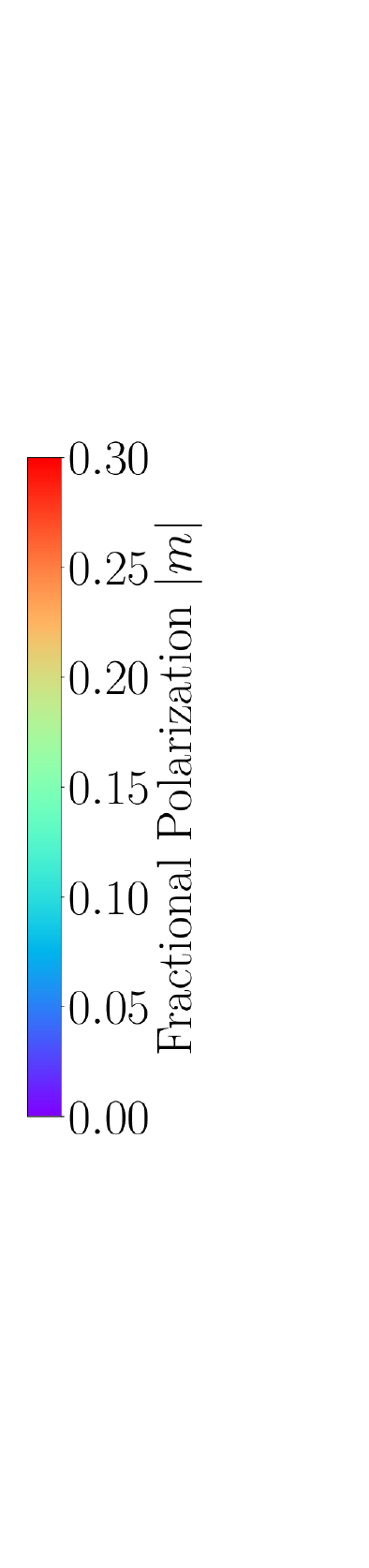}
    \caption{Linear polarization images of NRAO\,530 produced using {\tt DIFMAP}, {\tt eht-imaging}, {\tt DMC}, and {\tt Themis}. The total intensity is shown in grayscale with black contours indicating 10, 25, 50 and 75\% of the peak LP intensity. Black dotted contours indicate 25, 50, and 75\% of the peak polarized intensity. The ticks show the orientation of the EVPA, their length indicates linearly polarized intensity magnitude, and their color indicates fractional linear polarization. Cuts were made to omit all regions in the images where Stokes $\mathcal{I} <$ 10\% of the peak brightness and $\mathcal{P} <$ 10\% of the peak polarized brightness. The images are all displayed with a field of view of $128\,\mu$as, and all images are blurred to an equivalent resolution of $20\,\mu$as.
    \vspace{0.5eM}\\
}
    \label{fig:polimages}
\end{figure*}

\subsection{Linear Polarization Images}

\begin{figure}[t!]
    \centering
    \includegraphics[width=0.4\paperwidth]{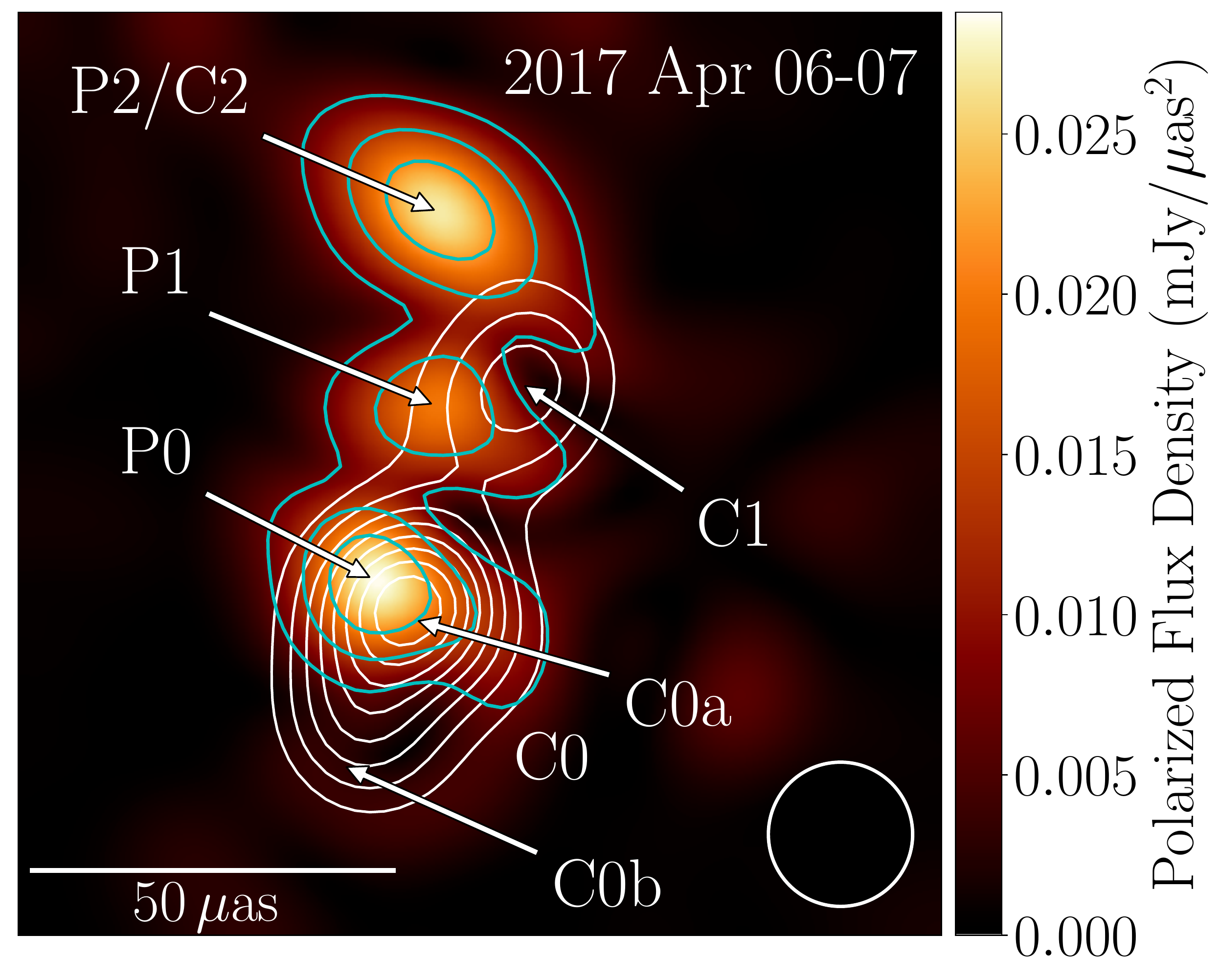}
    \caption{Schematic of the total-intensity (C0, C1, and C2) and linear polarization (P0, P1, and P2) components 
    in the EHT fiducial image of \nrao; white contours show the total intensity levels as in Figure~\ref{fig:1mmImage};
    color scale and cyan contours represent the polarized intensity of the method-averaged image. 
}
    \label{fig:PolAveImage}
\end{figure}

Figure~\ref{fig:polimages} presents LP  images of \nrao in 2017 April, obtained using a selection of methods. The absolute magnitude of linear polarization ($\sqrt{\mathcal{Q}^2 +\mathcal{U}^2}$ for LP Stokes image domain components $\mathcal{Q}$ and $\mathcal{U}$) is indicated with white contours and length of the electric vector position angle (EVPA) ticks, while the EVPA tick colors represent fractional LP magnitude. The Faraday rotation toward \nrao is entirely negligible at 230\,GHz \citep{Goddi2021}, hence observed EVPA can be directly interpreted as an intrinsic source property. The grayscale background image and black contours correspond to the total intensity. Similarly as in Section \ref{sec:fiducial_images_I}, in Figure \ref{fig:PolAveImage} we also present the method-averaged image, where the background image and cyan contours represent the LP magnitude, while the white contours describe the total intensity image map. Similarly as in case of the total intensity images, we adopted images shown in Figures~\ref{fig:polimages}-\ref{fig:PolAveImage} to a common field of view of $128\,\mu$as. The contribution of the emission outside of this region is consistent with the image noise. 

Figure \ref{fig:PolAveImage} serves the purpose of identifying the total intensity and LP components. As for the total intensity images, the LP images show similar structure across all algorithms, although there are some differences in the  polarization properties of the jet features. The most consistent among the methods is the polarized feature P0 near the core, which has a slight shift to the north-east with respect to the total intensity peak C0 in the \texttt{DIFMAP}, \texttt{eht-imaging}, and \texttt{DMC} images. In the \texttt{DMC} image the polarized feature in the south-east part of the C1 knot is most prominent, compared to images produced by the other methods. The \texttt{DIFMAP} and \texttt{DMC} images show a polarized feature, P2, associated with the diffuse northeastern total intensity component C2, which has the highest degree of polarization. A hint of this feature can be seen in the \texttt{eht-imaging} image, while the \texttt{Themis} image possesses a prominent polarized feature P1 that is rather a combination of polarization that can be associated with knot C1 and the northeastern component. Discrepancies seen in the \texttt{Themis} image may be related to the spline raster resolution limitations imposed by the MCMC algorithm. We expect that the method-averaged LP image, shown in Figure \ref{fig:PolAveImage}, should reveal the most reliable image domain polarization features. 

\subsection{Linear Polarization Component Modeling}
\label{sec:modeling_polar}

\begin{table*}
    \centering
    \caption{Parameters of polarized components}
    \centering
    \begin{center}
    \begin{tabularx}{0.62\paperwidth}{lccccccc}
    \hline 
    \hline 
         Knot &  $R$ ($\mu$as) & $\Theta$ (deg)  & $a$ ($\mu$as) & $Q$ (mJy) & $U$ (mJy) & $m$ (\%) & EVPA (deg) \\
         \hline
         (1)&(2)&(3)&(4)&(5)&(6)&(7)&(8)\\
       \hline 
        \multicolumn{8}{c}{{\tt DIFMAP} (Gaussian modeling)}\\
        \hline
         P0 & 5.7$\pm$2.1 &47$\pm$32 & 15.3$\pm$0.7 &9.8$\pm$4.0&14.8$\pm$5.0&5.2$\pm$1.3&28$\pm$7  \\
         P1 & 26.6$\pm$1.8&1$\pm$9 & 16.8$\pm$1.1 &3.0$\pm$4.0&$-$7.7$\pm$5.0&6.7$\pm$2.5&$-$34$\pm$11 \\
         P2 & 55.8$\pm$1.0&6$\pm$5 & 14.0$\pm$1.0 &$-$8.4$\pm$4.0&13.8$\pm$5.0&18.1$\pm$4.2&61$\pm$12\\
          Map & \nodata&\nodata&\nodata&$-$6.1$\pm$4.0&24.7$\pm$5.0&1.7$\pm$0.6&52$\pm$10\\
          ALMA & \nodata&\nodata&\nodata& $-8.6$ & $36.8$ &2.35$\pm$0.03&51.6$\pm$0.4\\
         \hline
        \multicolumn{8}{c}{{\tt eht-imaging} (Gaussian modeling)}\\
         \hline
          P0 & 0.0 & \nodata & 10.6 & 9.5 & 10.1 & 6.6 &  24.4 \\
          P1 & 28.5 & -21 & 15.8 & 11.8 & $-8.8$ & 13.6 &  $-18.4$ \\
          P2 & 44.3 & -24 & 42.8$\times$13.6 & $-6.9$ & 13.0 & 15.2 &  59.0 \\
          Map & \nodata & \nodata & \nodata & $13.6$ & 17.2 & 4.1 &  25.8 \\
         \hline
          \multicolumn{8}{c}{{\tt eht-imaging} (image domain, blur 15 $\mu$as)}\\
         \hline
          P0 & 7.3 & 68 & 11.5 & 7.6 & 0.5 & 8.0 &  2 \\
          P1 & 31.1 & 12 & 10.4 & 5.6 & $-0.5$ & 22 &  $-3$ \\
          P2 & 54.5 & 7 & 20.4 & $-9.8$ & 25.2 & 58 &  56 \\
          Map & \nodata & \nodata & \nodata & $-11.2$ & 15.5 & 2.7 &  63 \\
         \hline
         \hline 
    \end{tabularx}\\
    \end{center}
 {NOTES -- 1) values presented for the {\it Map} region are obtained by integrating modes of Stokes $\mathcal{I}$, $\mathcal{Q}$, and $\mathcal{U}$ in the entire VLBI field of view; 2) parameters presented for ALMA are results obtained by \cite{Goddi2021} on 2017 Apr 6.
  The columns are as follows: (1) - designation of knot; (2) - distance from the core in $\mu$as; (3) - position angle with respect to the core in degree; (4) - size in $\mu$as; (5) - Stokes Q value in mJy; (6) - Stokes U value in mJy; (7) - degree of polarization  in percentage; (8) - position angle of polarization in degree.
}  
   \label{tab:polmod}    
\end{table*}

The method-averaged polarized intensity image presented in Figure~\ref{fig:PolAveImage} can be described with three polarized features designated as P0, P1, and P2. Similarly as in Section~\ref{sec:modeling_total_int}, we have modeled Stokes $\mathcal{Q}$ and $\mathcal{U}$ visibilities in {\tt DIFMAP} using the task {\it modelfit}, in {\tt eht-imaging} Gaussian component geometric modeling framework, and we also used image-domain feature extraction methods to characterize the {\tt eht-imaging} results. Table~\ref{tab:polmod} gives parameters of the polarized components, which are as follows: 1) polarized component; 2) distance between the center of the component and the total intensity peak; 3) position angle of the component with respect to the center; 4)  FWHM size of the component; 5) Stokes $\mathcal{Q}$ parameter; 6) Stokes $\mathcal{U}$ parameter; 7) fractional degree of polarization; and 8) component EVPA.  Since $\mathcal{Q}$ and $\mathcal{U}$ data were modeled separately in {\tt DIFMAP}, the distance, position angle, and size of a component are calculated as averages over the $\mathcal{Q}$ and $\mathcal{U}$ values, and standard deviations of the averages are given as uncertainties. The uncertainties in $\mathcal{Q}$ and $\mathcal{U}$ flux intensity are equal to the root mean square (rms) of the $\mathcal{Q}$ and $\mathcal{U}$ map noise. Based on the uncertainties in $R$, $\Theta$, and $a$ values given in Table~\ref{tab:polmod}, there is very good agreement in the position and size of polarized components between $\mathcal{Q}$ and $\mathcal{U}$ images, which supports the robustness of the {\tt DIFMAP} modeling. We identify polarized components P0, P1, and P2 with the total intensity components C0, C1, and C2, respectively, with parameters of the latter given in Table~\ref{tab:fluxmod}. The distances of the associated total and polarized intensity knots agree within 1.5\,$\sigma$ uncertainties, and their sizes are almost identical, except the P2/C2 pair in {\tt DIFMAP} modeling where the polarized component P2 occupies only a fraction (less than a half) of knot C2. However, the position angles of P0 and P1 show some differences relative to those of C0 and C1, respectively, which cause the centers of the polarized components to be placed at the eastern edge of the corresponding total intensity components. The other EHT \sgra calibrator,  J1924-2914, does not exhibit similar effect \citepalias{Issaoun2022}, despite the fact that the two sources were calibrated in the same way and the same D-terms were employed to correct for the instrumental leakage. Hence, we conclude that these shifts are intrinsic to \nrao rather than caused by instrumental effects.

The LP image in Figure~\ref{fig:PolAveImage} shows polarized component P2, which we associate with total intensity knot C2, detected with Gaussian component models. However, due to its diffuse nature, the surface brightness of the knot is comparable with the noise level, so that C2 is indicated as a marginal feature in Figure~\ref{fig:1mmImage}, yet the knot has the highest degree of polarization in the jet. As mentioned above, in the case of the {\tt DIFMAP} model polarized emission of C2, knot P2, occupies only a part of the total intensity feature. If we fix the size and position of a total intensity component, C2, equal to those of P2, and search for the total flux density corresponding to the region of P2, this will yield a low value of $\sim$27~mJy, which implies the fractional polarization in the P2 region as high as $\sim$60\%. A similarly high degree of polarization is obtained in the case of {\tt eht-imaging} image domain analysis. Therefore, the P2 region should have an almost uniform magnetic field, which implies synchrotron emission fractional polarization of $m_{\rm max} = (\alpha+1)/(\alpha+5/3)\times100$=75\% \citep{Pacholczyk70}, assuming a spectral index for P2 of $\alpha\sim1$. One of possible explanations why the knot is prominent in polarization but not in the total intensity is that this feature could be caused by shear, ordering the field, but not by shocks that would make it bright in total intensity. 
For the geometric modeling of $\mathcal{Q}$ and $\mathcal{U}$ in {\tt eht-imaging} we assumed that each total intensity Gaussian component corresponds to an LP component with a constant EVPA and fractional polarization, obtaining a lower fractional polarization of C2/P2, which is nevertheless significantly larger than that of the core component P0.

Image domain feature extraction with {\tt eht-imaging} was based on masking disjointed image regions by LP flux density following the procedures used in \citetalias{Issaoun2022}. We obtained three components that could be identified with P0, P1, and P2 from Figure~\ref{fig:PolAveImage}. The results given in Table \ref{tab:polmod} are mostly in qualitative, if not quantitative, agreement between the methods, although the parameters agree within the 1.5~$\sigma$ uncertainties given for values obtained by {\tt DIFMAP}. The differences that are present can be attributed to the low polarized intensity in the jet relative to the polarized intensity noise and the systematic uncertainties specific to different imaging methods. 

Visual inspection of Figure~\ref{fig:polimages} suggests that EVPAs in the P0/C0 region lie between 20$^\circ$ -- 35$^\circ$ in all polarization images, except the {\tt eht-imaging} image, which is also consistent with the result of the $\mathcal{Q}$ and $\mathcal{U}$ map modeling in {\tt DIFMAP} and {\tt eht-imaging} Gaussian modeling. The {\tt eht-imaging} image shows the EVPA to be closer to 0$^\circ$, which can be explained by the higher level of polarization noise in the image. All images show polarized emission in the P1 region, with the EVPA close to the jet direction, while the EVPA of P2 is perpendicular to the inner jet position angle as seen in the {\tt DIFMAP} and {\tt DMC} images in Figure~\ref{fig:polimages} and determined by modeling. 

We have summed all CLEAN modes within the {\tt DIFMAP} total intensity image in Stokes parameter $\mathcal{I}$ (including the flux density of the BG component, see Table~\ref{tab:fluxmod}), and Stokes parameters $\mathcal{Q}$, and $\mathcal{U}$. This has resulted in an integrated total intensity flux of 1.5~Jy. Polarization parameters integrated over the sub-mas EHT image field of view are listed in Table~\ref{tab:polmod} under the component name  ``Map''. These parameters agree reasonably well with those obtained by \cite{Goddi2021} for \nrao at 221~GHz using ALMA data on 2017 Apr 6 and 7. There is very little variability of the flux density or polarization parameters between the two days, with flux densities reported by \cite{Goddi2021} equal to 1.61$\pm$0.16~Jy (Apr 6) and 1.57$\pm$0.16~Jy (Apr 7), and polarization parameters obtained on 2017 Apr~6 reproduced in Table~\ref{tab:polmod} under the component name ``ALMA''. Including Stokes parameters $\mathcal{Q}$ and $\mathcal{U}$ of the P2 knot in the calculation of Map values is essential for a good agreement with the polarization measurements obtained by \cite{Goddi2021}. 

\subsection{Constraints on Circular Polarization}

The circular polarization of \nrao at 230\,GHz observed by ALMA (but unresolved) is consistent with zero \citep{Goddi2021}. The high resolution of the EHT observation can be favourable for a detection of the circular polarization in the jet of \nrao, which is important for determining the plasma composition \citep{Wardle1998}. To find whether there is a statistically significant signature of spatially resolved circular polarization in the EHT VLBI data, we employed the same procedure as \citetalias{Issaoun2022}, based on exploring the posterior distributions of the \texttt{DMC} fits to the observational data. Using 1000 images drawn from the posterior distribution, and evaluating the mean and standard deviation of circular polarization in each reconstructed pixel, we find no pixel with a Stokes $\mathcal{V}$ detection of significance larger than 1.4\,$\sigma$. Hence, we conclude that we did not find statistically significant circular polarization resolved with the EHT resolution of $\sim 20\,\mu$as. For comparison, the same procedure applied to the linear polarization maps confirms detection at the $\sim 7\,\sigma$ level. Because of large image reconstruction uncertainties, this statement does not translate to a robust upper limit on the fractional circular polarization. 
\section{Discussion}
\label{sec:discussion}
The high-resolution total and polarized intensity images of the quasar \nrao at 230~GHz obtained during the EHT campaign in 2017 April indicate a jet direction of $\sim-$28$^\circ$ on $\mu$as scales. This jet direction is different from those seen at 86~GHz (PA$\sim$10$^\circ$, \citealt{Issaoun2019}) and 43~GHz (PA$\sim-$3$^\circ$, \citealt{Weaver2022}), implying a wiggling of the jet with the distance from the core. In fact, the PA of component C2/P2 (see Table \ref{tab:polmod}) agrees with that of the jet direction at 86~GHz within the uncertainties, which can be a sign of jet curvature already at this location. If that is indeed the case, the EVPA of the P2 component could be more aligned with the local direction of the bending jet, which supports the origin of the emission in the shear. The curvature could be caused by an imbalance between the pressure inside and outside the jet resulting in the development of instabilities in the flow, or it could be connected with jet precession. An analysis of the jet structure from sub-parsec to kiloparsec scales using multi-wavelength VLBI images of \nrao contemporaneous with the 230\,GHz EHT image will be presented in Lisakov et al. (in prep.). 

The brightest feature in the source, C0, which we associate with the VLBI core at 1.3~mm, is complex and consists of two components, C0a and C0b. Complex structure in the core was also revealed in the EHT images of the quasar 3C\,279 \citep{Kim2020}. Variability of such core structures can be expected on short timescales with quasi-periodic oscillations produced by
kink instabilities in the case of a magnetically dominated jet \citep{Dong2020}. The intrinsic brightness temperatures of both C0 components are lower than the equipartition brightness temperature (see Table~\ref{tab:fluxmod}), even if the Doppler factor of the jet on $\mu$as scales is lower than that found on mas scales. A possible explanation is that the energy density of the jet on $\mu$as scales is dominated by the magnetic field. 

\cite{Homan2006} have analyzed brightness temperatures of VLBI cores at 15~GHz of a large sample of AGN. They have found that, in a quiescent state, cores have a narrow range of intrinsic brightness temperatures close to the value expected for equipartition, while during outbursts the characteristic intrinsic brightness temperature significantly exceeds $T_{\rm eq}$. The authors have estimated that in active states the energy in radiating particles exceeds the energy in the magnetic field by a factor of $\sim$10$^5$. This implies that somewhere between the 230~GHz and 15~GHz cores there is a region where a magnetically dominated jet changes its state and reaches equipartition conditions. Moreover, \cite{J07} suggested, based on quasi-simultaneous multi-epoch observations at 43~GHz (VLBA), 86~GHz (BIMA\footnote{The Berkeley-Illinois-Maryland Array- Association (BIMA) operated at 3~mm and 1.3mm until 2004}), and 230/345~GHz (JCMT), that this region is located between the 86~GHz and 43~GHz VLBI cores, while \cite{Lee2013} has found that 86~GHz VLBI cores in a sample of compact radio sources observed with Global Millimeter VLBI Array have a characteristic intrinsic brightness temperature lower than the equipartition temperature. The observed low brightness temperature of C0 somewhat supports this scenario. However, analysis of the simultaneous VLBI observations at 230 and 86~GHz is needed to draw firm conclusions. 

The polarized features in the jet are shifted with respect to the total intensity components with which they are associated. The shifts are almost transverse to the inner jet direction -- the jet direction is from southeast to northwest, while the shifts are to the east. This feature appears robust in the image reconstructions. One viable interpretation of this phenomenon involves a difference in ordering of the magnetic field transverse to the jet, with the magnetic field on the eastern part of the jet being more ordered than that of the western part. This can be understood if the magnetic field in the jet consists of two components -- ordered and turbulent, with the turbulent component being more pronounced on the western side of the jet, while a helical magnetic field can represent the ordered component. Indeed, a helical structure of the magnetic field is supported by
the EVPA behavior in different polarized components. The EVPA undergoes significant changes along $\sim$60~$\mu$as of the jet length: it goes from being oblique in the core to aligning with the jet direction in P1 $\sim$30~$\mu$as to perpendicular to the inner jet direction in P2 $\sim$55~$\mu$as from the core. 
In addition, a higher polarization at the western side can be
attributed to velocity shear between the jet boundary and the ambient
medium being stronger on the western side, which is consistent with
the explanation of the nature of component P2/C2. However, the EVPA
behavior is more aligned with a helical magnetic field, although we do
not have a sufficient resolution to resolve the EVPA structure across the jet.
Recent work by \cite{Pushkarev2022} analyzed the polarization properties in a large sample of AGN observed with the VLBA at 15~GHz within the MOJAVE survey presented transverse fractional polarization slices beyond the core region for 307 sources. The authors have found that the majority of the sources (including \nrao) show a clear increase in fractional polarization towards the jet edges, with an asymmetric profile. They suggest that the observed patterns of polarization can be explained by helical magnetic fields with different pitch angles and different geometric and kinematic parameters of the jet, which is surrounded by a sheath. It therefore appears that a helical magnetic field structure persists in the quasar jet from sub-parsec to hundred-parsec scales.

\begin{figure}[t!]
    \centering
    \includegraphics[width=0.4\paperwidth, trim={0cm 5cm 0 2cm},clip]{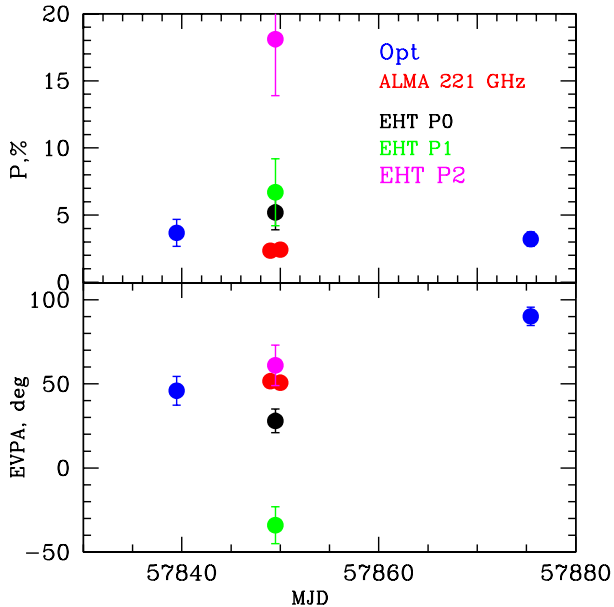}
    \caption{{\it Top:} Degree of linear polarization at optical R-band (blue), ALMA at 221~GHz (red), and EHT
    knots P0 (black), P1 (green), and P2 (magenta);
    {\it Bottom:} Position angle of polarization with the same color coding; parameters of knots are shown according to {\tt DIFMAP} model. }
    \label{fig:optpol}
\end{figure}

 In addition to EHT data, we have obtained optical polarization observations of \nrao several days before and after the EHT campaign. Figure~\ref{fig:optpol} shows polarization measurements in the optical R-band, at 221 GHz for the arcsec resolution of ALMA \citep{Goddi2021}, and those of the EHT resolved components at 230~GHz. The optical observations were carried out with the 1.8~m Perkins telescope (Flagstaff, AZ) using the PRISM camera. The description of data reduction can be found in \cite{J10}. 
 When comparing to the optical and ALMA values, the best agreement across wavelengths is found with the P2 component, indicating dominance of the magnetic field direction perpendicular to the $\mu$as scale jet direction in the source. This suggests that the magnetic field direction in the optical emission region is similar to that of P2, which has the highest degree of polarization in the jet.

\section{Summary}
\label{sec:conclusions}
Using the 2017 EHT observations we have obtained the first total and polarized intensity images of the quasar \nrao at 230 GHz, with an unprecedented angular resolution of $\sim$20\,${\mu}$as. We have not detected statistically significant variability of the source structure on a timescale of three days, and hence we have combined the data over epochs and frequencies to generate final images. We have used a variety of methods to construct the total and linearly polarized intensity images and averaged the final reconstructions produced by different methods to obtain the fiducial total and polarized intensity maps. 

On ${\mu}$as scales the quasar has a bright core of $\sim$300~mJy, with a 
jet extending up to 60\,${\mu}$as along a PA${\sim}-$28$^\circ$. The core has sub-structure
consisting of two features, C0a and C0b. We associate the 1.3~mm VLBI core with the brightest knot, C0a,
while C0b, located at a PA$\sim$+155$^\circ$ and $\sim$~3 times weaker than C0a can be a more opaque region of the core. The core is polarized, with a degree of polarization of $\sim$5$-$8\%.
The extended jet includes two features, C1/P1 and C2/P2, with mutually orthogonal directions of polarization, 
parallel and perpendicular to the jet, respectively, suggestive of a helical magnetic field
structure. The centers of the polarized knots P0 and P1 are shifted slightly transverse to the jet with respect 
to the centers of the corresponding total intensity knots C0 and C1. This can be caused either by a more ordered
magnetic field on the eastern side of the jet, connected with a helical structure of the magnetic field, or a stronger 
interaction between the jet and the external medium on this side. 

Feature C2 has the highest degree of polarization in the jet, $\sim20\%$. However, polarized knot P2, which occupies only a fraction of the total area of C2, has a peak polarization of $\sim60\%$. The latter implies a nearly uniform magnetic field in the P2 region. In fact, the optical polarization measurement nearest in time to the EHT campaign has an EVPA similar to that in P2. \nrao is a bright $\gamma$-ray source, and the origin of high-energy emission in blazars continues to be highly debated. We can speculate that the P2 region, with a very uniform magnetic field, might be a location where non-thermal optical and high energy emission originates. Future multi-epoch and multi-wavelength observations of the source simultaneous with EHT campaigns can provide a significant advance toward understanding the location of the high-energy dissipation zone in blazar jets.

\software{ \texttt{DIFMAP} \citep{Difmap}, \texttt{Matplotlib} \citep{matplotlib2007}, \texttt{DiFX} \citep{Deller2011}, \texttt{NumPy} \citep{numpy2011}, \texttt{eht-imaging} \citep{Chael_2016}, \texttt{PolConvert} \citep{Marti_2016}, \texttt{SMILI} \citep{Akiyama_2017a}, \texttt{EHT-HOPS} \citep{Blackburn_2019}, \texttt{Themis} \citep{Broderick2020}, \texttt{DMC} \citep{Dom2021} }



\section*{Acknowledgments}

The Event Horizon Telescope Collaboration thanks the following
organizations and programs: the Academia Sinica; the Academy
of Finland (projects 274477, 284495, 312496, 315721); the Agencia Nacional de Investigaci\'{o}n 
y Desarrollo (ANID), Chile via NCN$19\_058$ (TITANs) and Fondecyt 1221421, the Alexander
von Humboldt Stiftung; an Alfred P. Sloan Research Fellowship;
Allegro, the European ALMA Regional Centre node in the Netherlands, the NL astronomy
research network NOVA and the astronomy institutes of the University of Amsterdam, Leiden University and Radboud University;
the ALMA North America Development Fund; the Black Hole Initiative, which is funded by grants from the John Templeton Foundation and the Gordon 
and Betty Moore Foundation (although the opinions expressed in this work are those of the author(s) 
and do not necessarily reflect the views of these Foundations);
Chandra DD7-18089X and TM6-17006X; the China Scholarship
Council; China Postdoctoral Science Foundation fellowship (2020M671266); Consejo Nacional de Ciencia y Tecnolog\'{\i}a (CONACYT,
Mexico, projects  U0004-246083, U0004-259839, F0003-272050, M0037-279006, F0003-281692,
104497, 275201, 263356);
the Consejer\'{i}a de Econom\'{i}a, Conocimiento, 
Empresas y Universidad 
of the Junta de Andaluc\'{i}a (grant P18-FR-1769), the Consejo Superior de Investigaciones 
Cient\'{i}ficas (grant 2019AEP112);
the Delaney Family via the Delaney Family John A.
Wheeler Chair at Perimeter Institute; Direcci\'{o}n General
de Asuntos del Personal Acad\'{e}mico-Universidad
Nacional Aut\'{o}noma de M\'{e}xico (DGAPA-UNAM,
projects IN112417 and IN112820); 
the Dutch Organization for Scientific Research (NWO) VICI award
(grant 639.043.513) and grant OCENW.KLEIN.113; the Dutch National Supercomputers, Cartesius and Snellius  
(NWO Grant 2021.013); 
the EACOA Fellowship awarded by the East Asia Core
Observatories Association, which consists of the Academia Sinica Institute of Astronomy and
Astrophysics, the National Astronomical Observatory of Japan, Center for Astronomical Mega-Science,
Chinese Academy of Sciences, and the Korea Astronomy and Space Science Institute; 
the European Research Council (ERC) Synergy
Grant ``BlackHoleCam: Imaging the Event Horizon
of Black Holes" (grant 610058); 
the European Union Horizon 2020
research and innovation programme under grant agreements
RadioNet (No 730562) and 
M2FINDERS (No 101018682); the Horizon ERC Grants 2021 programme under grant agreement No. 101040021;
the Generalitat
Valenciana postdoctoral grant APOSTD/2018/177 and
GenT Program (project CIDEGENT/2018/021); MICINN Research Project PID2019-108995GB-C22;
the European Research Council for advanced grant `JETSET: Launching, propagation and 
emission of relativistic jets from binary mergers and across mass scales' (Grant No. 884631); 
the Institute for Advanced Study; the Istituto Nazionale di Fisica
Nucleare (INFN) sezione di Napoli, iniziative specifiche
TEONGRAV; 
the International Max Planck Research
School for Astronomy and Astrophysics at the
Universities of Bonn and Cologne; 
DFG research grant ``Jet physics on horizon scales and beyond'' (Grant No. FR 4069/2-1);
Joint Columbia/Flatiron Postdoctoral Fellowship, 
research at the Flatiron Institute is supported by the Simons Foundation; 
the Japan Ministry of Education, Culture, Sports, Science and Technology (MEXT; grant JPMXP1020200109); the Japanese Government (Monbukagakusho:
MEXT) Scholarship; 
the Japan Society for the Promotion of Science (JSPS) Grant-in-Aid for JSPS
Research Fellowship (JP17J08829); the Joint Institute for Computational Fundamental Science, Japan; the Key Research
Program of Frontier Sciences, Chinese Academy of
Sciences (CAS, grants QYZDJ-SSW-SLH057, QYZDJSSW-SYS008, ZDBS-LY-SLH011); 
the Leverhulme Trust Early Career Research
Fellowship; the Max-Planck-Gesellschaft (MPG);
the Max Planck Partner Group of the MPG and the
CAS; the MEXT/JSPS KAKENHI (grants 18KK0090, JP21H01137,
JP18H03721, JP18K13594, 18K03709, JP19K14761, 18H01245, 25120007); the Malaysian Fundamental Research Grant Scheme (FRGS) FRGS/1/2019/STG02/UM/02/6; the MIT International Science
and Technology Initiatives (MISTI) Funds; 
the Ministry of Science and Technology (MOST) of Taiwan (103-2119-M-001-010-MY2, 105-2112-M-001-025-MY3, 105-2119-M-001-042, 106-2112-M-001-011, 106-2119-M-001-013, 106-2119-M-001-027, 106-2923-M-001-005, 107-2119-M-001-017, 107-2119-M-001-020, 107-2119-M-001-041, 107-2119-M-110-005, 107-2923-M-001-009, 108-2112-M-001-048, 108-2112-M-001-051, 108-2923-M-001-002, 109-2112-M-001-025, 109-2124-M-001-005, 109-2923-M-001-001, 110-2112-M-003-007-MY2, 110-2112-M-001-033, 110-2124-M-001-007, and 110-2923-M-001-001);
the Ministry of Education (MoE) of Taiwan Yushan Young Scholar Program;
the Physics Division, National Center for Theoretical Sciences of Taiwan;
the National Aeronautics and
Space Administration (NASA, Fermi Guest Investigator
grants 80NSSC20K1567 and 80NSSC22K1571, NASA Astrophysics Theory Program grant 80NSSC20K0527, NASA NuSTAR award 
80NSSC20K0645); 
NASA Hubble Fellowship 
grants HST-HF2-51431.001-A, HST-HF2-51482.001-A awarded 
by the Space Telescope Science Institute, which is operated by the Association of Universities for 
Research in Astronomy, Inc., for NASA, under contract NAS5-26555; 
the National Institute of Natural Sciences (NINS) of Japan; the National
Key Research and Development Program of China
(grant 2016YFA0400704, 2017YFA0402703, 2016YFA0400702); the National
Science Foundation (NSF, grants AST-0096454,
AST-0352953, AST-0521233, AST-0705062, AST-0905844, AST-0922984, AST-1126433, AST-1140030,
DGE-1144085, AST-1207704, AST-1207730, AST-1207752, MRI-1228509, OPP-1248097, AST-1310896, AST-1440254, 
AST-1555365, AST-1614868, AST-1615796, AST-1715061, AST-1716327,  AST-1716536, OISE-1743747, AST-1816420, AST-1935980, AST-2034306); 
NSF Astronomy and Astrophysics Postdoctoral Fellowship (AST-1903847); 
the Natural Science Foundation of China (grants 11650110427, 10625314, 11721303, 11725312, 11873028, 11933007, 11991052, 11991053, 12192220, 12192223); 
the Natural Sciences and Engineering Research Council of
Canada (NSERC, including a Discovery Grant and
the NSERC Alexander Graham Bell Canada Graduate
Scholarships-Doctoral Program); the National Youth
Thousand Talents Program of China; the National Research
Foundation of Korea (the Global PhD Fellowship
Grant: grants NRF-2015H1A2A1033752, the Korea Research Fellowship Program:
NRF-2015H1D3A1066561, Brain Pool Program: 2019H1D3A1A01102564, 
Basic Research Support Grant 2019R1F1A1059721, 2021R1A6A3A01086420, 2022R1C1C1005255); 
Netherlands Research School for Astronomy (NOVA) Virtual Institute of Accretion (VIA) postdoctoral fellowships; 
Onsala Space Observatory (OSO) national infrastructure, for the provisioning
of its facilities/observational support (OSO receives
funding through the Swedish Research Council under
grant 2017-00648);  the Perimeter Institute for Theoretical
Physics (research at Perimeter Institute is supported
by the Government of Canada through the Department
of Innovation, Science and Economic Development
and by the Province of Ontario through the
Ministry of Research, Innovation and Science); the Princeton Gravity Initiative; the Spanish Ministerio de Ciencia e Innovaci\'{o}n (grants PGC2018-098915-B-C21, AYA2016-80889-P,
PID2019-108995GB-C21, PID2020-117404GB-C21); 
the University of Pretoria for financial aid in the provision of the new 
Cluster Server nodes and SuperMicro (USA) for a SEEDING GRANT approved towards these 
nodes in 2020;
the Shanghai Pilot Program for Basic Research, Chinese Academy of Science, 
Shanghai Branch (JCYJ-SHFY-2021-013);
the State Agency for Research of the Spanish MCIU through
the ``Center of Excellence Severo Ochoa'' award for
the Instituto de Astrof\'{i}sica de Andaluc\'{i}a (SEV-2017-
0709); the Spinoza Prize SPI 78-409; the South African Research Chairs Initiative, through the 
South African Radio Astronomy Observatory (SARAO, grant ID 77948),  which is a facility of the National 
Research Foundation (NRF), an agency of the Department of Science and Innovation (DSI) of South Africa; 
the Toray Science Foundation; the Swedish Research Council (VR); 
the US Department
of Energy (USDOE) through the Los Alamos National
Laboratory (operated by Triad National Security,
LLC, for the National Nuclear Security Administration
of the USDOE (Contract 89233218CNA000001); and the YCAA Prize Postdoctoral Fellowship.

We thank
the staff at the participating observatories, correlation
centers, and institutions for their enthusiastic support.
This paper makes use of the following ALMA data:
ADS/JAO.ALMA\#2016.1.01154.V. ALMA is a partnership
of the European Southern Observatory (ESO;
Europe, representing its member states), NSF, and
National Institutes of Natural Sciences of Japan, together
with National Research Council (Canada), Ministry
of Science and Technology (MOST; Taiwan),
Academia Sinica Institute of Astronomy and Astrophysics
(ASIAA; Taiwan), and Korea Astronomy and
Space Science Institute (KASI; Republic of Korea), in
cooperation with the Republic of Chile. The Joint
ALMA Observatory is operated by ESO, Associated
Universities, Inc. (AUI)/NRAO, and the National Astronomical
Observatory of Japan (NAOJ). The NRAO
is a facility of the NSF operated under cooperative agreement
by AUI.
This research used resources of the Oak Ridge Leadership Computing Facility at the Oak Ridge National
Laboratory, which is supported by the Office of Science of the U.S. Department of Energy under Contract
No. DE-AC05-00OR22725. We also thank the Center for Computational Astrophysics, National Astronomical Observatory of Japan.
The computing cluster of Shanghai VLBI correlator supported by the Special Fund 
for Astronomy from the Ministry of Finance in China is acknowledged.
This work was supported by FAPESP (Fundacao de Amparo a Pesquisa do Estado de Sao Paulo) under grant 2021/01183-8.

APEX is a collaboration between the
Max-Planck-Institut f{\"u}r Radioastronomie (Germany),
ESO, and the Onsala Space Observatory (Sweden). The
SMA is a joint project between the SAO and ASIAA
and is funded by the Smithsonian Institution and the
Academia Sinica. The JCMT is operated by the East
Asian Observatory on behalf of the NAOJ, ASIAA, and
KASI, as well as the Ministry of Finance of China, Chinese
Academy of Sciences, and the National Key Research and Development
Program (No. 2017YFA0402700) of China
and Natural Science Foundation of China grant 11873028.
Additional funding support for the JCMT is provided by the Science
and Technologies Facility Council (UK) and participating
universities in the UK and Canada. 
The LMT is a project operated by the Instituto Nacional
de Astr\'{o}fisica, \'{O}ptica, y Electr\'{o}nica (Mexico) and the
University of Massachusetts at Amherst (USA). The
IRAM 30-m telescope on Pico Veleta, Spain is operated
by IRAM and supported by CNRS (Centre National de
la Recherche Scientifique, France), MPG (Max-Planck-Gesellschaft, Germany) 
and IGN (Instituto Geogr\'{a}fico
Nacional, Spain). The SMT is operated by the Arizona
Radio Observatory, a part of the Steward Observatory
of the University of Arizona, with financial support of
operations from the State of Arizona and financial support
for instrumentation development from the NSF.
Support for SPT participation in the EHT is provided by the National Science Foundation through award OPP-1852617 
to the University of Chicago. Partial support is also 
provided by the Kavli Institute of Cosmological Physics at the University of Chicago. The SPT hydrogen maser was 
provided on loan from the GLT, courtesy of ASIAA.

This work used the
Extreme Science and Engineering Discovery Environment
(XSEDE), supported by NSF grant ACI-1548562,
and CyVerse, supported by NSF grants DBI-0735191,
DBI-1265383, and DBI-1743442. XSEDE Stampede2 resource
at TACC was allocated through TG-AST170024
and TG-AST080026N. XSEDE JetStream resource at
PTI and TACC was allocated through AST170028.
This research is part of the Frontera computing project at the Texas Advanced 
Computing Center through the Frontera Large-Scale Community Partnerships allocation
AST20023. Frontera is made possible by National Science Foundation award OAC-1818253.
This research was carried out using resources provided by the Open Science Grid, 
which is supported by the National Science Foundation and the U.S. Department of 
Energy Office of Science. 
Additional work used ABACUS2.0, which is part of the eScience center at Southern Denmark University. 
Simulations were also performed on the SuperMUC cluster at the LRZ in Garching, 
on the LOEWE cluster in CSC in Frankfurt, on the HazelHen cluster at the HLRS in Stuttgart, 
and on the Pi2.0 and Siyuan Mark-I at Shanghai Jiao Tong University.
The computer resources of the Finnish IT Center for Science (CSC) and the Finnish Computing 
Competence Infrastructure (FCCI) project are acknowledged. This
research was enabled in part by support provided
by Compute Ontario (http://computeontario.ca), Calcul
Quebec (http://www.calculquebec.ca) and Compute
Canada (http://www.computecanada.ca). 

The EHTC has
received generous donations of FPGA chips from Xilinx
Inc., under the Xilinx University Program. The EHTC
has benefited from technology shared under open-source
license by the Collaboration for Astronomy Signal Processing
and Electronics Research (CASPER). The EHT
project is grateful to T4Science and Microsemi for their
assistance with Hydrogen Masers. This research has
made use of NASA's Astrophysics Data System. We
gratefully acknowledge the support provided by the extended
staff of the ALMA, both from the inception of
the ALMA Phasing Project through the observational
campaigns of 2017 and 2018. We would like to thank
A. Deller and W. Brisken for EHT-specific support with
the use of DiFX. We thank Martin Shepherd for the addition of extra features in the Difmap software 
that were used for the CLEAN imaging results presented in this paper.
We acknowledge the significance that
Maunakea, where the SMA and JCMT EHT stations
are located, has for the indigenous Hawaiian people.


\bibliography{bibliography.bib}        


\vfill\eject
\end{document}